\documentclass[prl,twocolumn,showpacs,amsmath,amssymb]{revtex4-1}

\usepackage{graphicx}
\usepackage{amsmath,amssymb,bm}
\usepackage{color}

\begin{document}

\title{
  Fermi liquid theory for nonlinear transport through a multilevel Anderson impurity
}

 \author{Yoshimichi Teratani}
 \affiliation{
 Department of Physics, Osaka City University, Sumiyoshi-ku, 
 Osaka 558-8585, Japan }

\author{Rui Sakano}
\affiliation{
The Institute for Solid State Physics, 
the University of Tokyo, Kashiwa, Chiba 277-8581, Japan
}

 \author{Akira Oguri}
 \affiliation{
 Department of Physics, Osaka City University, Sumiyoshi-ku, 
 Osaka 558-8585, Japan }
\affiliation{Nambu Yoichiro Institute of Theoretical and Experimental Physics, 
Osaka City University, Osaka 558-8585, Japan}

\date{\today}

\begin{abstract}
We present a microscopic Fermi-liquid view on the  
low-energy transport through  
an Anderson impurity with $N$ discrete levels,  at arbitrary electron filling $N_d$. 
It is applied to nonequilibrium current fluctuations,  
for which the two-quasiparticle collision integral 
and the three-body correlations that determine the quasiparticle energy shift 
 play  important roles. 
Using the numerical renormalization group up to $N=6$, 
we find that for strong interactions the three-body fluctuations 
are determined by a single parameter other than the Kondo energy scale 
 in a wide filling range  $1 \lesssim N_d \lesssim N-1$. 
It significantly affects  the current noise for $N>2$  
 and the behavior of noise in magnetic fields.  
\end{abstract}


\maketitle

{\it Introduction.---}
Highly correlated low-energy states of the Kondo systems 
show fascinating universal behavior \cite{HewsonBook},   
which can be described by a Fermi liquid (FL) theory  in zero dimension 
\cite{WilsonRMP,NozieresFermiLiquid,YamadaYosida4,ShibaKorringa,Yoshimori}. 
FL behaviors have been observed for the nonlinear current 
through quantum dots \cite{GrobisGoldhaber-Gordon,ScottNatelson}   
and also the current noise \cite{Heiblum,Delattre2009,KobayashiKondoShot,Ferrier2016}
which is now  one of the most important probes to explore quantum states. 
Furthermore,  in addition to the spin,   
 internal degrees of freedom  such as {\it orbital, flavor,} etc., 
 bring an interesting variety in the Kondo effect, occurring   
in a carbon nanotube \cite{RMP-Kouwenhoven,Ferrier2016} 
and novel quantum systems such as 
 ultracold atomic gases \cite{TakahashiColdGasKondo} 
 and quark matters \cite{QCD_Kondo}.

Transport properties of the local FL  have successfully been 
described by the renormalized quasiparticles and their collisions 
due to the residual interaction,   
especially at the {\it symmetric point\/} where   
 both the particle-hole (PH) and time-reversal (TR) symmetries 
are present \cite{Hershfield1,ao2001PRB,Aligia,Munoz}.  
These symmetries are broken in real systems 
by  external fields, such as a gate voltage and a magnetic field.   
In this case, a single quasiparticle captures the 
quadratic dependence on frequency $\omega$,  temperature $T$, 
and bias voltage $V$ not only through the well-investigated {\it damping rate\/} 
 but also through the {\it energy shift\/}. 
It has recently been clarified that the quadratic {\it energy shift\/} is determined 
by the three-body correlations between the impurity electrons 
\cite{MoraMocaVonDelftZarand,FilipponeMocaWeichselbaumVonDelftMora,ao2017_1_PRL,ao2017_3_PRB_addendum,ao2017_3_PRB_erratum_2}. 
 It shows that the three-body correlations are    
 essential parameters for describing the FL transport.

Despite its importance, the current noise 
 \cite{Hershfield2,GogolinKomnikPRL,Sela2006,SelaMalecki,MoraSUnKondoII,SakanoFujiiOguri,KarkiMoraVonDelfKiselex}  has been 
still less elucidated away from the {\it symmetric point\/}. 
A major milestone was achieved by Mora {\it et al\/} 
\cite{MoraMocaVonDelftZarand}, 
who  have  extended Nozi{\`e}res 
phenomenological FL theory   \cite{NozieresFermiLiquid} 
 to give the formula of the nonlinear noise   
for a PH asymmetric single-orbital Anderson model  
at zero magnetic field.
Further investigation, however, is required to clarify 
the physics of nonequilibrium  fluctuations 
in the Kondo systems with various internal degrees of freedom.

In this letter,  we give a microscopic view   
on  the low-energy transport  through a multilevel Anderson impurity 
for a wide range of electron fillings $N_d$. 
It is described in terms of   {\it five\/} FL parameters, 
which can be calculated using 
the numerical renormalization group (NRG) 
\cite{WilsonRMP}  up to  $N=6$. 
We find that for strong interactions 
 the three-body 
correlations for $N$ degenerate levels are determined by  a single  parameter  
over a wide filling range  $1 \lesssim N_d \lesssim N-1$,  
which includes  the intermediate valence regions.  
We also provide a current-noise formula for the FL,  taking into account  
all the two-quasiparticle collision processes  \cite{LandauPhysicalKinetics,HaugJauho}.  
It satisfies a Ward identity \cite{YamadaYosida4,ShibaKorringa,Yoshimori} 
for the Keldysh vertex function,  
and resolves  an essential problem of the current conservation 
of the correlated electrons under a  nonequilibrium  condition 
\cite{Hershfield1,Hershfield2}.   
We also calculate the nonlinear  noise using the NRG,  
 and demonstrate that the internal degrees of freedom give a wide variety 
to the filling dependence.
We also examine the effect of a magnetic field that breaks the TR symmetry,
and show that the noise of a spin-1/2 quantum dot exhibits 
a universal Kondo scaling behavior.

{\it Model.---}
We consider an $N$-level Anderson impurity 
coupled to two leads on the left ($L$)  and right ($R$): 
\begin{align}
& 
\! 
\mathcal{H} \, =   \   
 \sum_{\sigma=1}^N
 \epsilon_{d\sigma}^{}\, n_{d\sigma}  
\,
+    \sum_{\lambda=L,R} \sum_{\sigma=1}^N   v_{\lambda}^{}
 \left( \psi_{\lambda\sigma}^\dag d_{\sigma}^{} + 
  d_{\sigma}^{\dag} \psi_{\lambda\sigma}^{} \right) 
\nonumber 
\\ 
& 
\quad  \ 
 + \sum_{\lambda=L,R} \sum_{\sigma=1}^N 
\int_{-D}^D  \! d\epsilon\,  \epsilon\, 
 c^{\dagger}_{\epsilon \lambda \sigma} c_{\epsilon \lambda \sigma}^{}
\,+ 
\frac{U}{2} \sum_{\sigma \neq \sigma'}^{} 
n_{d\sigma}^{} n_{d\sigma'}^{} .
 \label{eq:H}
\end{align}
 $d^{\dag}_{\sigma}$ creates 
 an impurity electron with energy $\epsilon_{d\sigma}$, 
  $n_{d\sigma} \equiv d^{\dag}_{\sigma} d^{}_{\sigma}$, and   
  $U$  the Coulomb repulsion. 
Conduction electrons are normalized as 
$\{ c^{\phantom{\dagger}}_{\epsilon\lambda\sigma}, 
c^{\dagger}_{\epsilon'\lambda'\sigma'}
\} = \delta_{\lambda\lambda'} \,\delta_{\sigma\sigma'}   
\delta(\epsilon-\epsilon')$. 
 The coupling $v_{\lambda}^{}$ between 
$\psi^{}_{\lambda\sigma} \equiv  \int_{-D}^D d\epsilon \sqrt{\rho_c^{}}  
\, c^{\phantom{\dagger}}_{\epsilon\lambda \sigma}$
and $d^{\dag}_{\sigma}$  
yields  a resonance of the width  
 $\Delta \equiv \Gamma_L + \Gamma_R$,  with  
 $\Gamma_{\lambda} = \pi \rho_c^{} v_{\lambda}^2$, 
 $\,\, \rho_c^{}=1/(2D)$, and  $D$ the half band width.

In this work, we study the nonlinear current noise  
 \cite{Hershfield2} 
\begin{align}
\!\!\!\!
S_\mathrm{noise}^\mathrm{QD}  = \!
\int_{-\infty}^{\infty} \!\!  dt\, 
\left\langle 
\delta \widehat{J}(t) \, \delta \widehat{J}(0) 
+\delta \widehat{J}(0) \, \delta \widehat{J}(t)
\right\rangle_{V}^{} \;.
\label{eq:S_noise}
\end{align}
Here,  $\delta \widehat{J}(t)  \equiv  \widehat{J}(t) 
- \langle  \widehat{J}(0) \rangle_{V}^{}$ 
is the current fluctuation operator through the quantum dot 
\footnote{
 ${\protect \widehat{J}}
 \equiv    
(\Gamma_L  {\protect \widehat{J}}_{R} 
+\Gamma_R {\protect \widehat{J}}_{L})/(\Gamma_L+\Gamma_R)$,    
with ${\protect \widehat{J}}$ 
the current flowing between the dot and lead on $\lambda$ ($=L,R$) side.
},
and $\langle \cdots \rangle_{V}^{}$ is   
the Keldysh steady-state average defined 
at finite bias voltages $eV \equiv \mu_L-\mu_R$ 
with $\mu_{\lambda}$ the chemical potential for $\lambda=L,R$  
\footnote{See supplemental material for details.}. 
The average  current  
$J \equiv \langle  \widehat{J}(0) \rangle_{V}^{}$ 
is given by  \cite{Hershfield1},  
\begin{align}
& 
J \,=\, 
 \frac{e}{h}
\sum_{\sigma}  
\int_{-\infty}^{\infty} \!\! d\omega\,  
\bigl[\,f_L(\omega)-f_R(\omega) \,\bigr]\, 
  \mathcal{T}_{\sigma}(\omega)  \,.
\label{eq:current_formula}
\end{align}
Here,  $f_{\lambda}(\omega) \equiv [e^{(\omega-\mu_\lambda)/T}+1]^{-1}$     
the Fermi function, 
$\mathcal{T}_{\sigma}(\omega) \equiv     
   - \frac{4\Gamma_L \Gamma_R}{\Gamma_L +\Gamma_R} 
 \, \mathrm{Im} \,G_{\sigma}^{r}(\omega) $ 
the transmission probability, and   
$G_{\sigma}^{r}(\omega)
=[\,\omega -\epsilon_{d\sigma}^{} +i\Delta \,
- \Sigma_{\sigma}^{r}(\omega) \,]^{-1}$ 
the retarded Green's function with 
$\Sigma_{\sigma}^{r}(\omega)$  
the self-energy.   
From this  $\mathcal{T}_{\sigma}(\omega)$, we can also deduce  
 the thermal conductivity $\kappa_\mathrm{QD}^{}$
\footnote{
$
\kappa_\mathrm{QD}^{} \, \equiv \,    
\frac{1}{h T}
\bigl[  \, 
\sum_{\sigma}
 \mathcal{L}_{2,\sigma}^{\mathrm{QD}} 
\ - \  
{ \bigl(
\sum_{\sigma}
\mathcal{L}_{1,\sigma}^{\mathrm{QD}}
\bigr)^2}/
{
\bigl(\sum_{\sigma}
 \mathcal{L}_{0,\sigma}^{\mathrm{QD}}
\bigr)
} 
\, \bigr]  
$,   
and
 $\mathcal{L}_{n,\sigma}^\mathrm{QD}  =   
 \int_{-\infty}^{\infty}  
 d\omega\,  \omega^n  (-\frac{\partial f}{\partial \omega})\, 
  \mathcal{T}_{\sigma}(\omega)$ for $n=0,1,2$.
} 
for the heat current  $J_Q^{}= -  \kappa_\mathrm{QD}^{}\, \delta T$,        
 induced by the temperature difference  $\delta T$ 
between the two leads \cite{CostiZlatic2010}.

{\it Fermi-liquid parameters.---}
We investigate low-energy transport up to next leading order.
To this end, we  expand  $\Sigma_{\sigma}^{r}(\omega)$  
up to terms of order $\omega^2$, $T^2$, and $(eV)^2$ for general $N$,  
 extending the latest FL description for spin $1/2$ case 
\cite{ao2017_1_PRL,ao2017_3_PRB_addendum}. 
The expansion coefficients play an important role as the  FL parameters.

The phase shift   $\delta_{\sigma}^{}  \equiv 
\cot^{-1} (\epsilon_{d\sigma}^{*}/\Delta)$ 
is a parameter of primary importance, 
 with  $\epsilon_{d\sigma}^{*} 
\equiv \epsilon_{d\sigma}^{} + 
\left. \Sigma_{\sigma}^{r}(0) \right|_{T= eV=0}^{}$
 the effective  impurity level. 
 It determines 
the occupation number $\langle n_{d\sigma} \rangle =\delta_{\sigma}^{}/\pi$,
and 
the density of states 
$\rho_{d\sigma}^{} \equiv {\sin^2 \delta_{\sigma}^{}}/{(\pi\Delta)}$.
The renormalization factor is given by
the first derivative 
 $z_{\sigma}^{} \equiv
\bigl[1 -  
\frac{\partial  \Sigma_{\sigma}^{r}(\omega)}{\partial \omega}
|_{\omega = 0}\bigr]^{-1}$,  defined 
at $T=eV=0$. 
It is also related to the static susceptibility   
 $\chi_{\sigma_1\sigma_2}^{} \equiv  
 \int_0^{1/T}  \!\! d \tau \, 
\left\langle  \delta n_{d\sigma_1}(\tau)\,\delta  n_{d\sigma_2}\right
\rangle$,  
  as $\chi_{\sigma\sigma}^{} 
\xrightarrow{T\to 0}\rho_{d\sigma}^{}/z_{\sigma}^{}$,  
with $\delta n_{d\sigma} \equiv n_{d\sigma}-
\langle n_{d\sigma} \rangle$ 
\cite{YamadaYosida4,ShibaKorringa,Yoshimori}.   
The second derivative is a complex number, 
the imaginary part of which 
corresponds to 
 the single-quasiparticle  {\it damping rate\/} 
of order $\omega^2$, $T^2$, and $(eV)^2$ 
\cite{Hershfield1,ao2001PRB}.
The real part corresponds to the quadratic {\it  energy shift\/} 
that is determined by  the nonlinear susceptibility defined at equilibrium  
\cite{ao2017_1_PRL,ao2017_3_PRB_addendum}:
\begin{align}
\chi_{\sigma_1\sigma_2\sigma_3}^{[3]} 
\! 
\equiv  & 
- \!
\int_{0}^{1/T} \!\!\!\! d\tau_3 \!\! 
\int_{0}^{1/T} \!\!\!\! d\tau_2\, 
\langle T_\tau 
\delta n_{d\sigma_3} (\tau_3) \,
\delta n_{d\sigma_2} (\tau_2) \,
\delta n_{d\sigma_1}
\rangle \,,
\nonumber 
\end{align}
with $T_\tau$ the imaginary-time ordering operator.
It can also be written as   
$ \chi_{\sigma_1\sigma_2\sigma_3}^{[3]} =
{\partial \chi_{\sigma_1\sigma_2}^{}}/{\partial \epsilon_{d\sigma_3}^{}}$, 
and contributes to the transport when the PH or TR symmetry is broken.  



\begin{widetext}

\begin{table}[bt]
\caption{Coefficients $C$'s introduced in Eq.\ 
\eqref{eq:FL_noise}.
$W$'s and $\Theta$'s represent the two- and three-body contributions, respectively. 
}
\begin{tabular}{l|l} 
\hline \hline
\ $C_{S}^{} 
\ =  \,   
\frac{\pi^2}{192} 
\left[\, W_S^{}  
-
\cos 2\delta\,
\Bigl\{
 \Theta_\mathrm{I}^{} 
+
3 (N-1) \Theta_\mathrm{II}^{} 
\Bigr\} 
\,\right]$ \ \ \ \ \  \ \ 
& \ \ \ 
$W_S^{} 
 \   \equiv  \, 
\cos 4 \delta\,
+ 
\Bigl[\,
 4+5\cos 4 \delta  + 
\frac{3}{2}\bigl(1- \cos 4\delta \bigr)\,(N-2)
\,\Bigr] (N-1)\left(R-1\right)^2 $
\rule{0cm}{0.45cm}
\\
\ $
C_{V}^{} 
\ =  \, 
\frac{\pi^2}{64}
 \,\bigl[
\,
W_{V}^{} 
\,+ \,
\Theta_\mathrm{I}^{}
+
3(N-1)\,
\Theta_\mathrm{II}^{}
\, \bigr]$ 
& \  \ \ 
$W_{V}^{} 
\  \equiv \, 
-
\left[\,
1
+ 
5(N-1) \left(R-1\right)^2
\,\right] \cos 2 \delta $ 
\rule{0cm}{0.45cm}
\\
\ $C_{T}^{} 
\ =   \, 
\frac{\pi^2}{48}
 \,\bigl[\,
W_{T}^{} 
\,+ \,
\Theta_\mathrm{I}^{}
+
(N-1)\,\Theta_\mathrm{II}^{}
 \,\bigr] \quad$
& \ \ \  
$W_{T}^{} 
\ \equiv \, 
-
\left[\,
1
+ 
2(N-1) \left(R-1\right)^2
\,\right] \cos 2 \delta $
\rule{0cm}{0.45cm}
\\
\ $C_{\kappa}^\mathrm{QD} 
=  \,   
\frac{7\pi^2}{80} 
\,\bigl[\,
W_{\kappa}^\mathrm{QD} 
\,+\, 
\Theta_\mathrm{I}^{} 
+
\frac{5}{21}(N-1)\, 
\Theta_\mathrm{II}^{} 
\,\bigr]$ 
  &  \ \ \  
$ W_{\kappa}^\mathrm{QD} 
  \equiv  
\frac{10 - 11\cos 2\delta}{21}
 - \frac{6}{7}(N-1)  \left( R  -1 \right)^2  \cos 2\delta$ 
\rule{0cm}{0.45cm}
\\
\hline
\hline
\end{tabular}
\label{tab:C_W_SUN}
\end{table}

\end{widetext}

{\it SU($N$) symmetric case.---}
In the case
at which the $N$  impurity levels 
are degenerate  $\epsilon_{d\sigma}^{}\equiv \epsilon_{d}^{}$, 
the linear susceptibility  $\chi_{\sigma\sigma'}^{}$ has 
only two independent components. The diagonal element determines 
the  energy scale  $T^* \equiv  1/(4\chi_{\sigma\sigma}^{})$, by which   
the  $T$-linear specific heat  is scaled as
  $\mathcal{C}_\mathrm{imp} = \frac{N \pi^2}{12}(T/T^*)$. 
It can also be identified as 
the Kondo temperature in the strong-coupling limit. 
The other one is  the off-diagonal element 
$\chi_{\sigma\sigma'}^{}$  for $\sigma \neq \sigma'$, 
which is related to the Wilson ratio   
  $R \equiv 1- \chi_{\sigma\sigma'}^{}/\chi_{\sigma\sigma}^{}$  
\cite{HewsonRPT2001}.  
Similarly, the nonlinear susceptibility  
has three independent components for $N\geq 3$: the diagonal element  
$\chi_{\sigma\sigma\sigma}^{[3]} $ and 
two off-diagonal ones,   
which can also  be expressed in the following form  
for  $\sigma \neq \sigma' \neq \sigma'' \neq \sigma$,  
\begin{align}
 -(N-1)\,\chi_{\sigma\sigma'\sigma'}^{[3]} 
\,=&  \ 
 \chi_{\sigma\sigma\sigma}^{[3]} 
- \frac{\partial \chi_{\sigma\sigma}^{}}{\partial  \epsilon_{d}^{}} 
\;, 
\label{eq:3body_multi_relation}
\\ 
\frac{(N-1)(N-2)}{2}\,\chi_{\sigma\sigma'\sigma''}^{[3]} 
= & \ 
\chi_{\sigma\sigma\sigma}^{[3]} 
  -\frac{\partial \chi_{\sigma\sigma}^{}}{\partial  \epsilon_{d}^{}} 
+\frac{N-1}{2}\,\frac{\partial \chi_{\sigma\sigma'}^{}}{\partial  \epsilon_{d}^{}} .
\nonumber 
 \end{align}

In this work, we obtain the low-energy expansion of 
 $S_\mathrm{noise}^\mathrm{QD}$, 
 $J$ and  $\kappa_\mathrm{QD}^{}$ 
up to next leading order,   specifically for symmetric junctions 
  $\Gamma_L=\Gamma_R$ and $\mu_L=-\mu_R =eV/2$:  
\begin{align}
&
\!\!\!
S_\mathrm{noise}^\mathrm{QD}
=    \frac{2Ne^2|eV| }{h} 
\left[\,
\frac{\sin^2 2\delta}{4} 
 + C_{S}^{} 
\left(\frac{eV}{T^*}\right)^2
 + \cdots
\right], 
\label{eq:FL_noise}
\\
&
\!\!\!\!
\frac{dJ}{dV} =  
\frac{Ne^2}{h} 
\left[\, 
\sin^2 \delta
- C_{T}^{} \left(\frac{\pi T}{T^*}\right)^2 
-   C_{V}^{} \left(\frac{eV}{T^*} \right)^2  
+ \cdots 
\right] , 
\nonumber
\\
&
\!\!\!
\kappa_\mathrm{QD}^{} \,= \,   
\frac{N\pi^2 T}{3 h}
\,  \left[\,
\sin^2 \delta
\,- 
 C_{\kappa}^\mathrm{QD}
\, 
\left( \frac{\pi T}{T^*}\right)^2
 + \cdots
\right] .
\nonumber 
\end{align}
The explicit expressions of the coefficients 
$C_{S}^{}$,  $C_{V}^{}$,  $C_{T}^{}$
and  $C_{\kappa}^\mathrm{QD}$ 
 are listed in table \ref{tab:C_W_SUN}. 
Each of these $C$'s  consists of two parts, 
denoted as $W$ and  $\Theta$.   
The  $W$-part 
represents  two-body contributions 
which can be described  in terms of $R$ and $\delta$.   
The $\Theta$-part represents dimensionless three-body contributions:  
\begin{align}
\!\!\!\!\!\!\! 
\Theta_\mathrm{I}^{} 
\,\equiv& \ 
\frac{\sin 2\delta}{2\pi}\,
\frac{\chi_{\sigma\sigma\sigma}^{[3]}}{\chi_{\sigma\sigma}^2}\,,
\qquad 
  \Theta_\mathrm{II}^{} 
\,\equiv \, 
\frac{\sin 2\delta}{2\pi}\,
\frac{\chi_{\sigma\sigma'\sigma'}^{[3]} }{\chi_{\sigma\sigma}^2}\,. 
\label{eq:Theta_definition}
\end{align}
Therefore,  the low-energy transport of  
 the SU($N$) Fermi liquid are determined completely by {\it five\/}  parameters:  
  $\delta$, $T^*$, $R$,  $\Theta_\mathrm{I}^{}$, 
and $\Theta_\mathrm{II}^{}$.   
These FL parameters can also be deduced experimentally  
through measurements of the coefficients $C$'s.
We note that another parameter  for three different levels,  
 $\Theta_\mathrm{III}^{} 
\equiv 
\frac{\sin 2\delta}{2\pi}\,
\frac{\chi_{\sigma\sigma'\sigma''}^{[3]} }{\chi_{\sigma\sigma}^2}$,   
 does {\it not\/} affect  $C$'s for symmetric junctions. 
Nevertheless,  it  contributes to the transport  for $N\geq 3$ 
when the tunneling couplings or the chemical potentials are asymmetric.

The  nonlinear noise  of the  Fermi liquid  
is determined not only by a single-quasiparticle excitation  
but also by two-quasiparticle collisions 
described by the Keldysh vertex corrections \cite{Hershfield2}.  
In this work, we calculate the vertex function up to order $eV$ \cite{Note2}, 
 extending  the diagrammatic approach of Yamada-Yosida 
\cite{YamadaYosida4,ShibaKorringa,Yoshimori}.     
Consequently, the collision contributions  $C_S^\mathrm{coll}$ 
and the single-quasiparticle ones  $C_S^\mathrm{qp}$ 
yield the nonlinear noise  
  $C_S^{} = C_S^\mathrm{qp} +  C_S^\mathrm{coll}$:       
\begin{align} 
 C_S^\mathrm{coll}  
   =  
\left[\,
\frac{ 7+5\cos 4 \delta}{2}  + 
\frac{3}{2}\,\bigl(1- \cos 4\delta\bigr)
\,(N-2)
\,\right] 
\frac{\widetilde{K}^2}{N-1}   
\nonumber 
\end{align}
with  $\widetilde{K} \equiv (N-1)(R-1)$. 
The second term in the bracket emerges through the collisions specific to   
 multilevel impurities for $N >2$, and it  vanishes in the SU(2) symmetric case or  
 the PH symmetric case  at which $\delta =\pi/2$.

\begin{figure}[t]

 \leavevmode
 \centering
\includegraphics[width=0.47\linewidth]{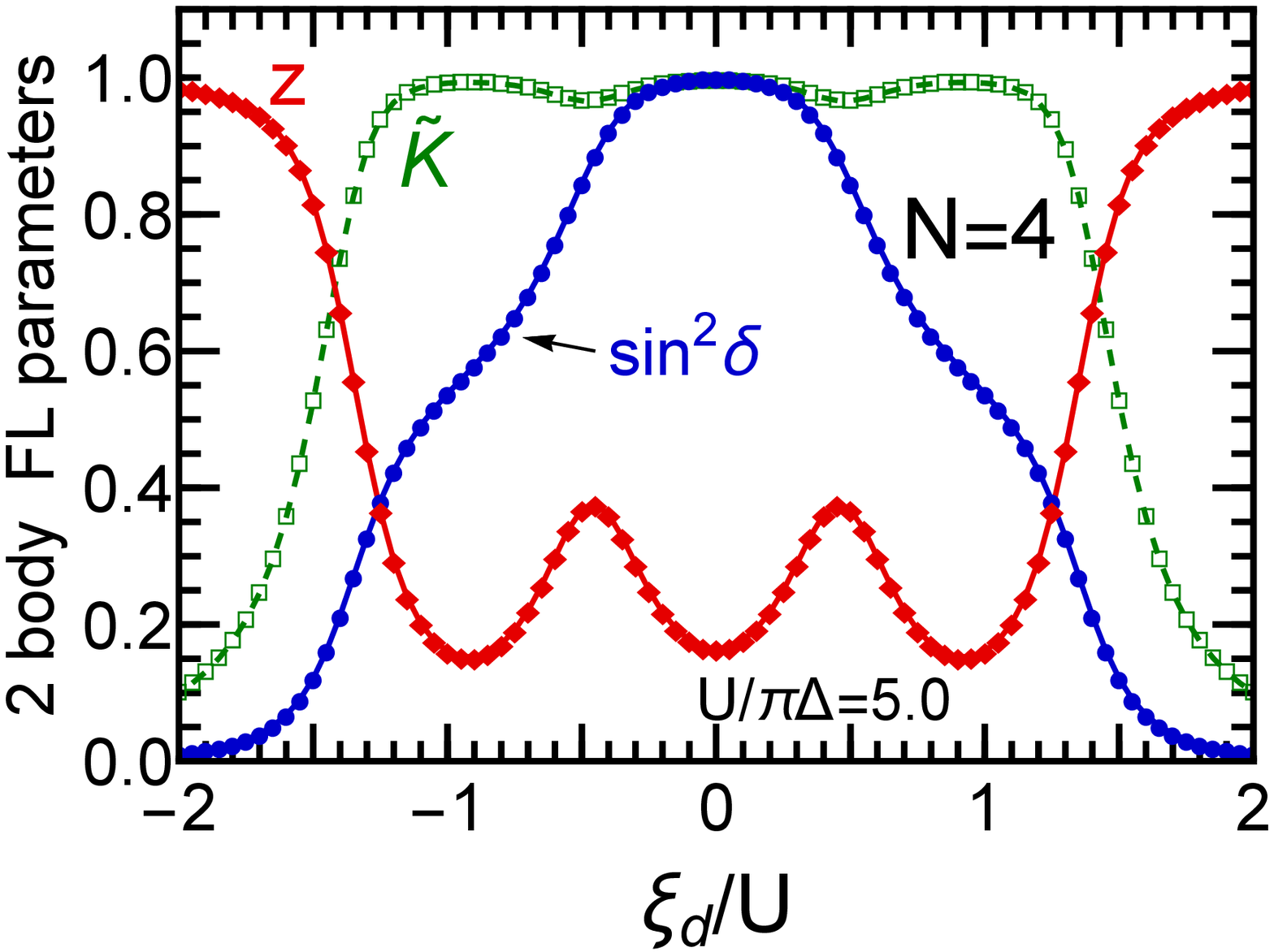}
 \hspace{0.01\linewidth} 
\includegraphics[width=0.47\linewidth]{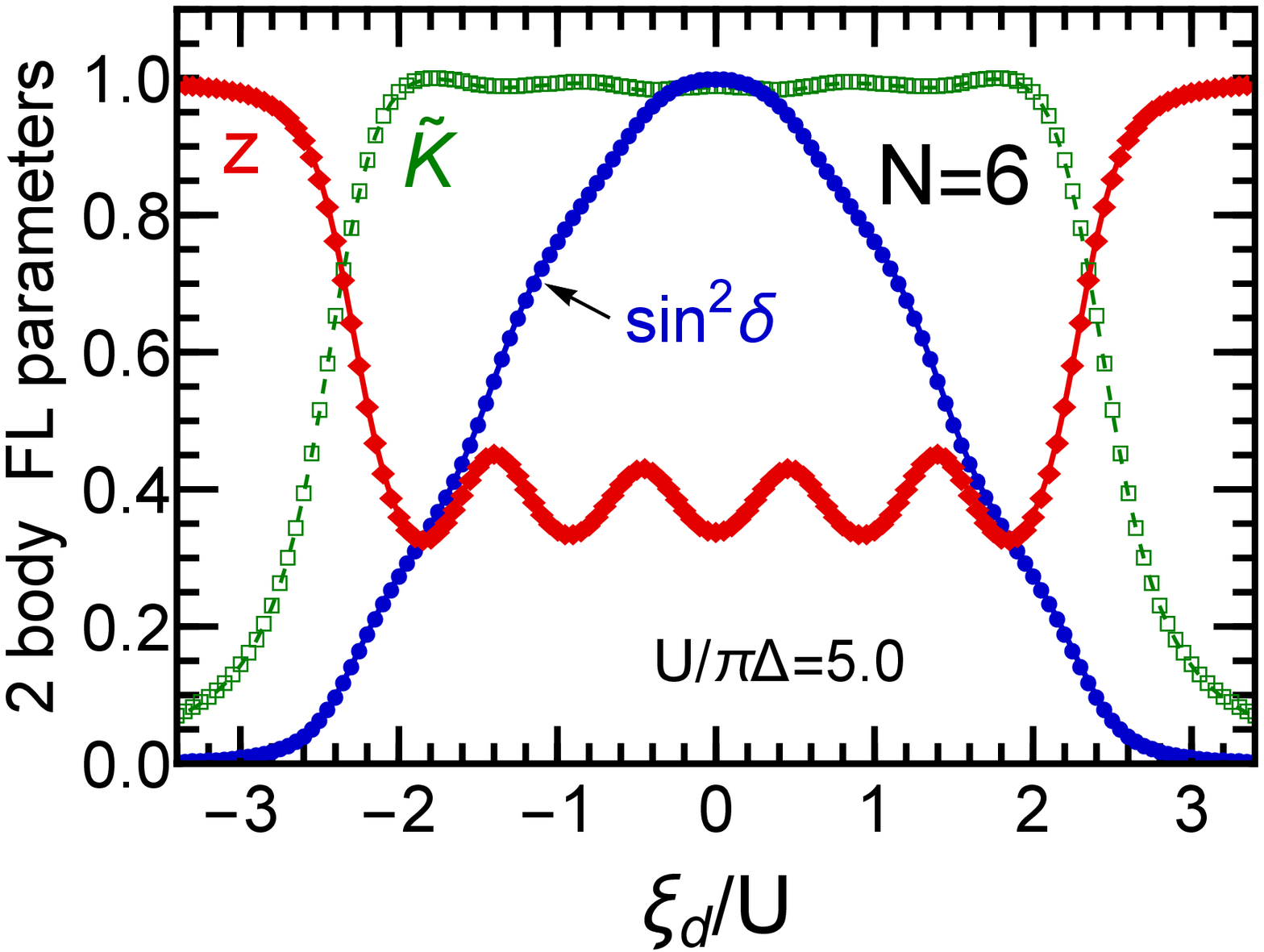}
\\
\includegraphics[width=0.47\linewidth]{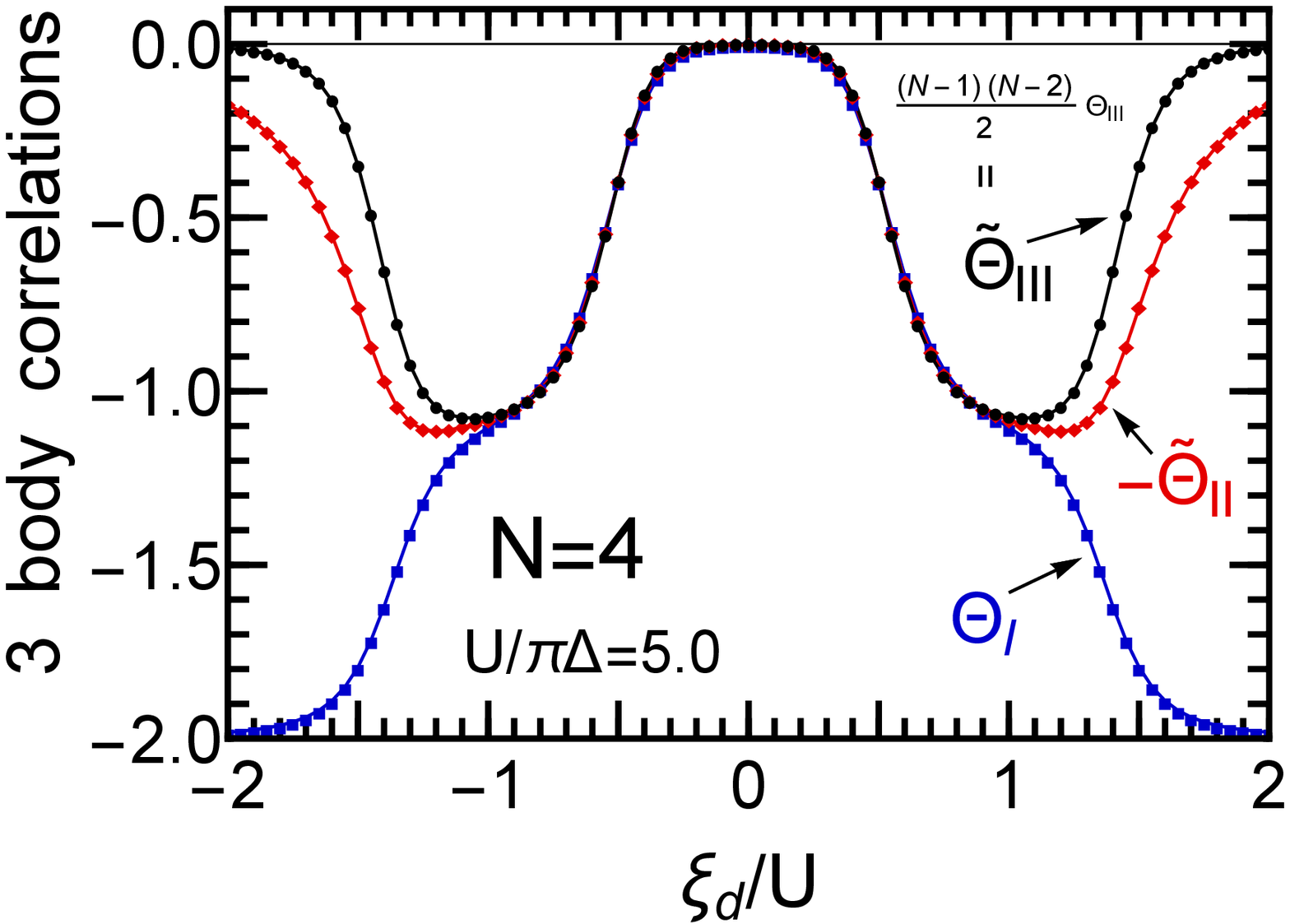}
 \hspace{0.01\linewidth} 
\includegraphics[width=0.47\linewidth]{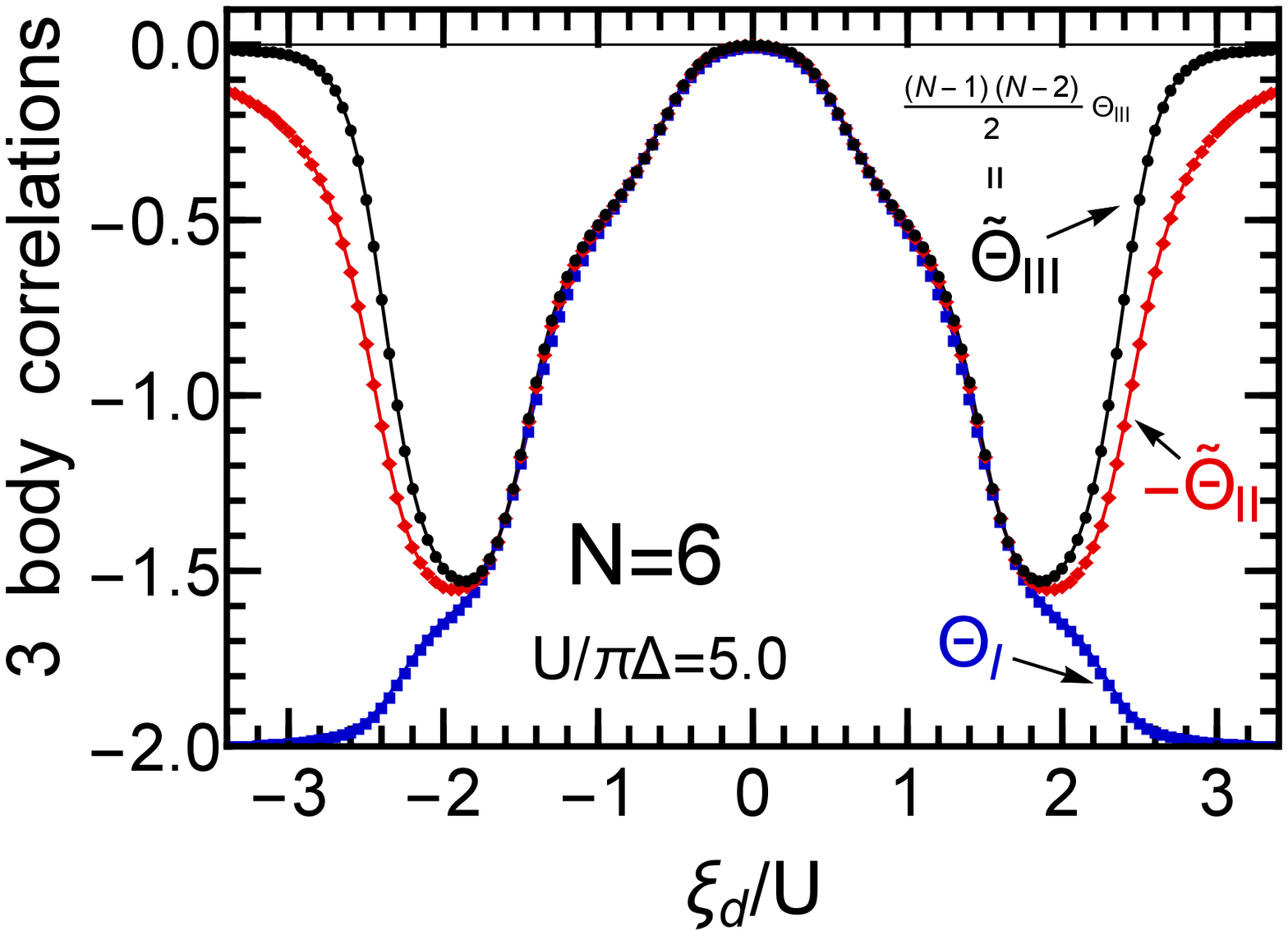}
\\
\includegraphics[width=0.47\linewidth]{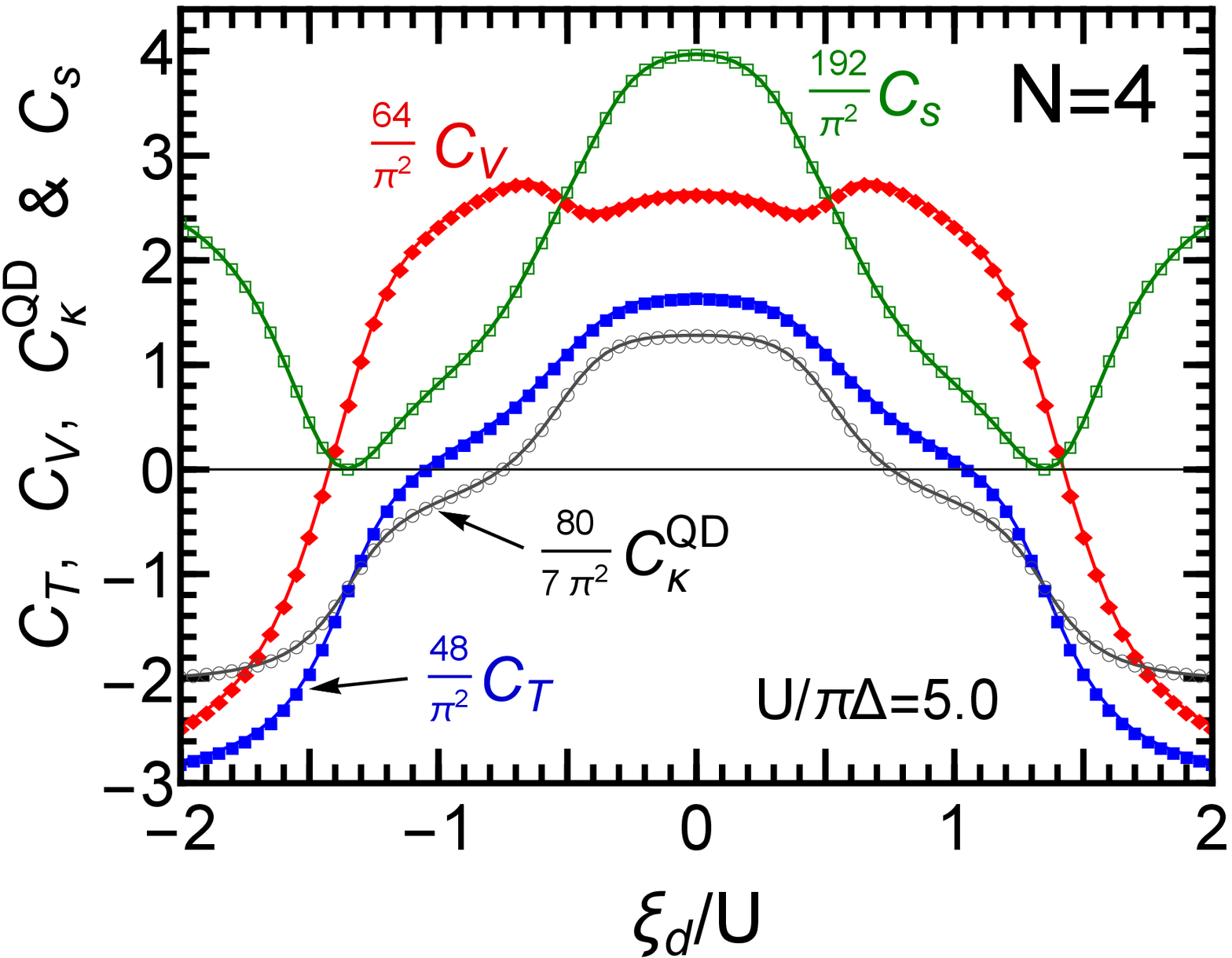}
 \hspace{0.01\linewidth} 
\includegraphics[width=0.47\linewidth]{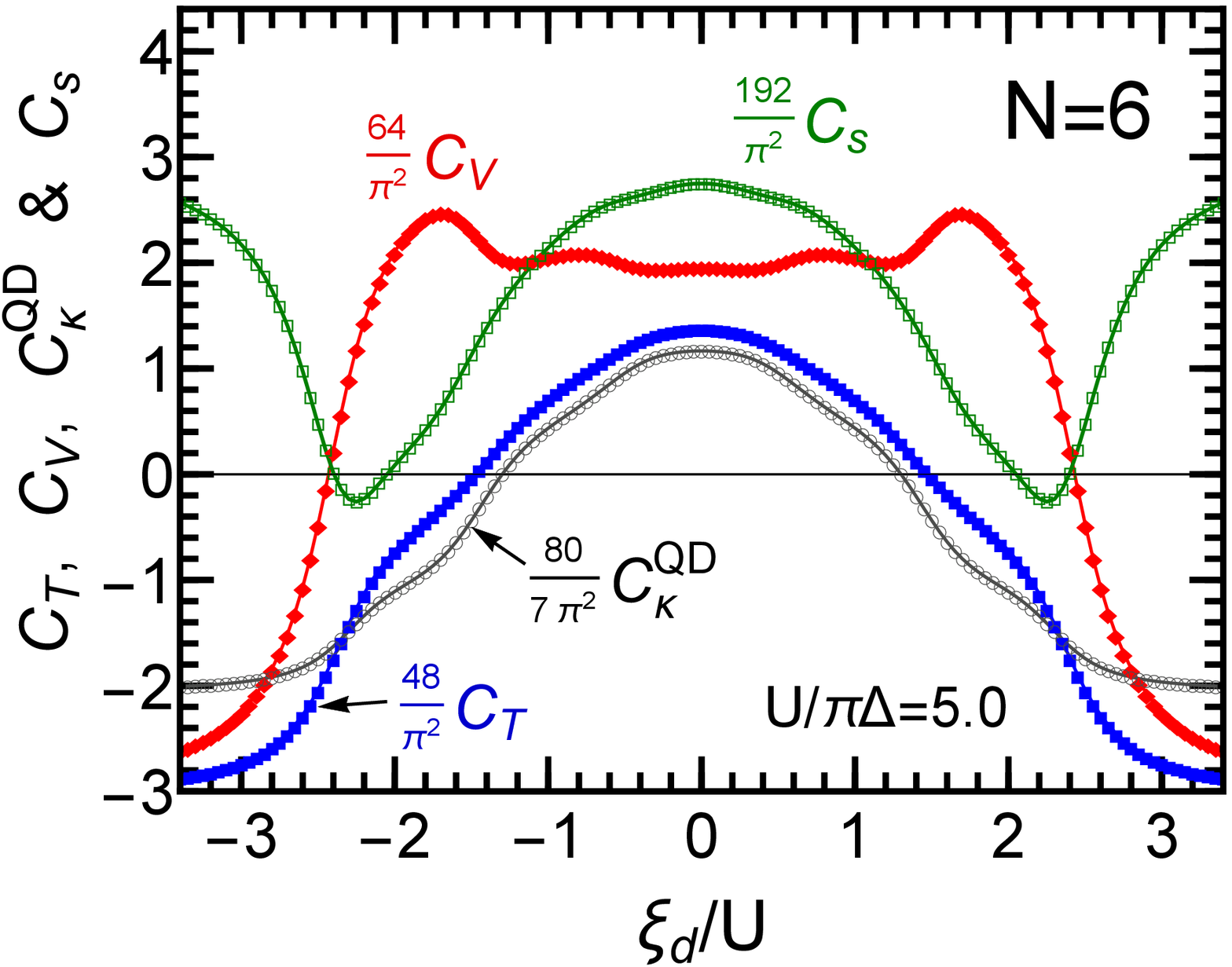}
\caption{
Fermi-liquid parameters for SU($N$) Anderson model  
are plotted vs  $\xi_{d}^{} \equiv \epsilon_d +(N-1)U/2$ 
for  $U/(\pi\Delta)=5.0$, 
 $N=4$  (left panels)  and  $N=6$ (right panels).  
Top panels:    
 $\sin^2 \delta$,  renormalization factor $z$,  and 
 $\widetilde{K}\equiv (N-1)(R-1)$. 
Middle  panels: 
$\Theta_\mathrm{I}^{}$, 
$\,\,-\widetilde{\Theta}_\mathrm{II}^{} 
\equiv  -(N-1) \Theta_\mathrm{II}^{}$,  and 
$\, \, \widetilde{\Theta}_\mathrm{III}^{} \equiv  \frac{(N-1)(N-2)}{2}
 \Theta_\mathrm{III}^{}$. 
Bottom  panels:  
 $\frac{48}{\pi^2} C_{T}^{}$,
$\frac{64}{\pi^2} C_{V}^{} $,
$\frac{80}{7\pi^2}C_{\kappa}^{\mathrm{QD}}$,  
and  $\frac{192}{\pi^2} C_{S}^{}$.  
}
\label{fig:W_Theta_SUN}
\end{figure}

{\it Filling dependence of the FL state.---} 
How does the FL state evolve   
as the number of levels  $N$ and their position $\epsilon_d^{}$ vary? 
As the electron configuration  
 $N_{d}^{}\equiv \sum_\sigma \langle n_{d\sigma}^{}\rangle$  
continuously varies with $\epsilon_d^{}$,  
a different class of the Kondo and valence-fluctuation states emerge 
 for multilevel systems $N>2$.
To our knowledge, however, the behavior of three-body correlations 
$\Theta$'s that determine the nonlinear transport 
has not been explored so much,   
whereas the two-body correlations have been well investigated  
for $N=4$ \cite{NishikawaCrowHewson2,SakanoFujiiOguri,ao2012}. 
In this work, 
we  calculate the FL parameters for $N=4, 6$ with the NRG,   
using the interleaved algorithm  particularly for $N=6$ \cite{Stadler2016}. 
To be specific, we choose the Coulomb interaction to be 
much larger than the hybridization energy scale: $U/(\pi \Delta) = 5.0$.  
The results are plotted vs  $\xi_{d}^{} \equiv \epsilon_d +(N-1)U/2$ 
in Fig.\  \ref{fig:W_Theta_SUN} for (left panels) $N=4$   and (right panels) $N=6$.

The top panels of Fig.\ \ref{fig:W_Theta_SUN} show the 
two-body correlations,
 relating to 
$\, \langle n_{d\sigma}^{}\rangle$, 
 $\chi_{\sigma\sigma}^{}$, and 
$\chi_{\sigma\sigma'}^{}$.
We see that  $\sin^2 \delta$, 
which determines $\mathcal{T}_{\sigma}(0)$ at $T=eV=0$, 
 shows a flat Kondo ridge of the unitary limit $\delta \simeq \pi/2$
near the PH symmetric point  $|\xi_{d}^{}| \lesssim U/2$ where 
the occupation number is almost locked at  $N_d^{}\simeq N/2$.     
The other Kondo ridges  
also emerge at $\xi_d^{}$ where $N_d^{}$ approaches an integer:   
 $\xi_{d}^{} \simeq \pm U$ for $N=4$, 
and also $\xi_{d}^{} \simeq \pm U, \pm 2U$ for $N=6$.

The renormalization factor $z$, which   
determines  the energy scale  $T^* = z\,\pi \Delta/(4\sin^2 \delta)$,  
 is also shown in the top panels. 
It is significantly suppressed over a wide range  
 of gate voltages  $|\xi_{d}^{}| \lesssim \frac{N-1}{2}U$,  
and  appears as a broad valley. 
 This valley  becomes shallow as $N$ increases,  
and vanishes in the  large $N$ limit \cite{ao2012}.  
Inside the valley, $z$ has  
minimums at  $\xi_{d}^{} \simeq  \frac{N-2M}{2} \, U$ 
for  $M = 1, 2, \ldots, N-1$, 
where the occupation number approaches an integer $N_d^{} = M$.
At these minimums,  the low-energy states 
can be described by the SU($N$) Kondo model 
 in the strong-coupling limit $U\gg \Delta$.    
We find that  $z$ is also suppressed  
 at local maximums corresponding to the intermediate valence states, 
 for both  $N=4$ and $6$. 
In  the top panels,  the rescaled Wilson ratio  $\widetilde{K}$ is also shown. 
It is  almost saturated to the  universal value $\widetilde{K} = 1$  
and its derivative  becomes very 
small  ${\partial \widetilde{K}}/{\partial \epsilon_{d}} \sim 0$ 
 in the whole region of  the broad valley  $1 \lesssim N_d^{}\lesssim N-1$.
It reveals the fact that not only the charge susceptibility  
$\chi_c^{} \equiv 
- 
{\partial \langle n_{d\sigma}^{}\rangle}/{\partial\epsilon_d^{}}
 =   \chi_{\sigma\sigma}^{} ( 1 - \widetilde{K})$     
but  its derivative ${\partial \chi_c^{}}/{\partial\epsilon_d^{}}$ 
is suppressed   in this region.

The three-body correlation $\Theta_\mathrm{I}^{}$ is plotted 
 in the middle panels of Fig.\  \ref{fig:W_Theta_SUN},  
together with the other two rescaled ones: 
$\,-\widetilde{\Theta}_\mathrm{II}^{} \equiv  -(N-1) \Theta_\mathrm{II}^{}$,  
and 
$\, \widetilde{\Theta}_\mathrm{III}^{} \equiv  \frac{(N-1)(N-2)}{2}
\Theta_\mathrm{III}^{}$. 
These  $\Theta$'s  also 
 show  plateau structures due to the Kondo effect 
 at the values of $\xi_d^{}$ corresponding to  integer $N_d^{}$, 
and almost 
vanish at  $|\xi_d| \lesssim  \frac{U}{2}$.
We find that these three parameters $\Theta_\mathrm{I}^{}$, 
$\,-\widetilde{\Theta}_\mathrm{II}^{}$, and 
$\widetilde{\Theta}_\mathrm{III}^{}$  
approach each other very closely 
over a wide gate-voltage range $|\xi_d| \lesssim  \frac{N-2}{2}\,U$,   
at which  $1 \lesssim N_d^{} \lesssim N-1$.  
This indicates that contributions of the diagonal element 
 $\chi_{\sigma\sigma\sigma}^{[3]}$ 
dominate the terms  in the right-hand side 
of Eq.\ \eqref{eq:3body_multi_relation};  
 i.e.\  
 $\chi_{\sigma\sigma\sigma}^{[3]}$  becomes much greater than 
 $\frac{\partial \chi_{\sigma\sigma}^{}}{\partial  \epsilon_{d}^{}}$ and 
 $\frac{N-1}{2}\,\frac{\partial \chi_{\sigma\sigma'}^{}}
{\partial  \epsilon_{d}^{}}$.
 It also reveals the fact  that  not only  $\partial \chi_c^{}/\partial \epsilon_d^{}$ 
but  also $\partial \chi_s^{}/\partial \epsilon_d^{}$, 
the derivative of the spin susceptibility 
$\chi_s^{}  \propto \chi_{\sigma\sigma}^{}-\chi_{\sigma\sigma'}^{}$, 
becomes much smaller than  $(T^*)^{-2}$.  
Thus,  for large $U$,    
the FL properties are characterized by {\it three\/} parameters   
 $\,\delta$,  $T^*$ and  $\Theta_\mathrm{I}^{}$ 
over the wide filling range  $1 \lesssim N_d^{} \lesssim N-1$.  
%
%
Outside this region,   
the  $\Theta$'s approach the noninteracting values:  
 $\Theta_\mathrm{I}^{} \to -2$,  
and the other two vanish as  $|\xi_d| \to \infty$.

\begin{figure}[t]

 \leavevmode
 \centering
\includegraphics[width=0.465\linewidth]{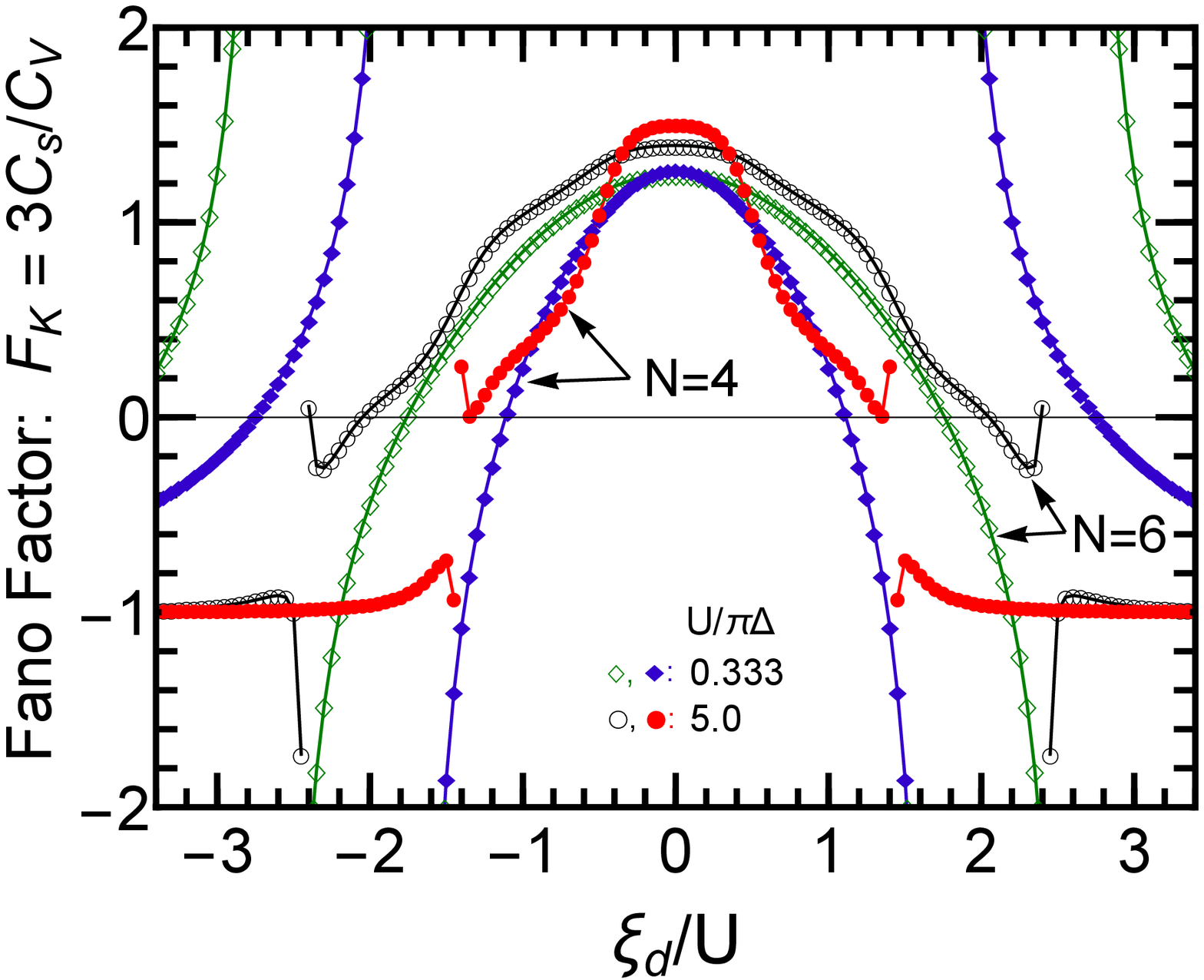}
\includegraphics[width=0.505\linewidth]{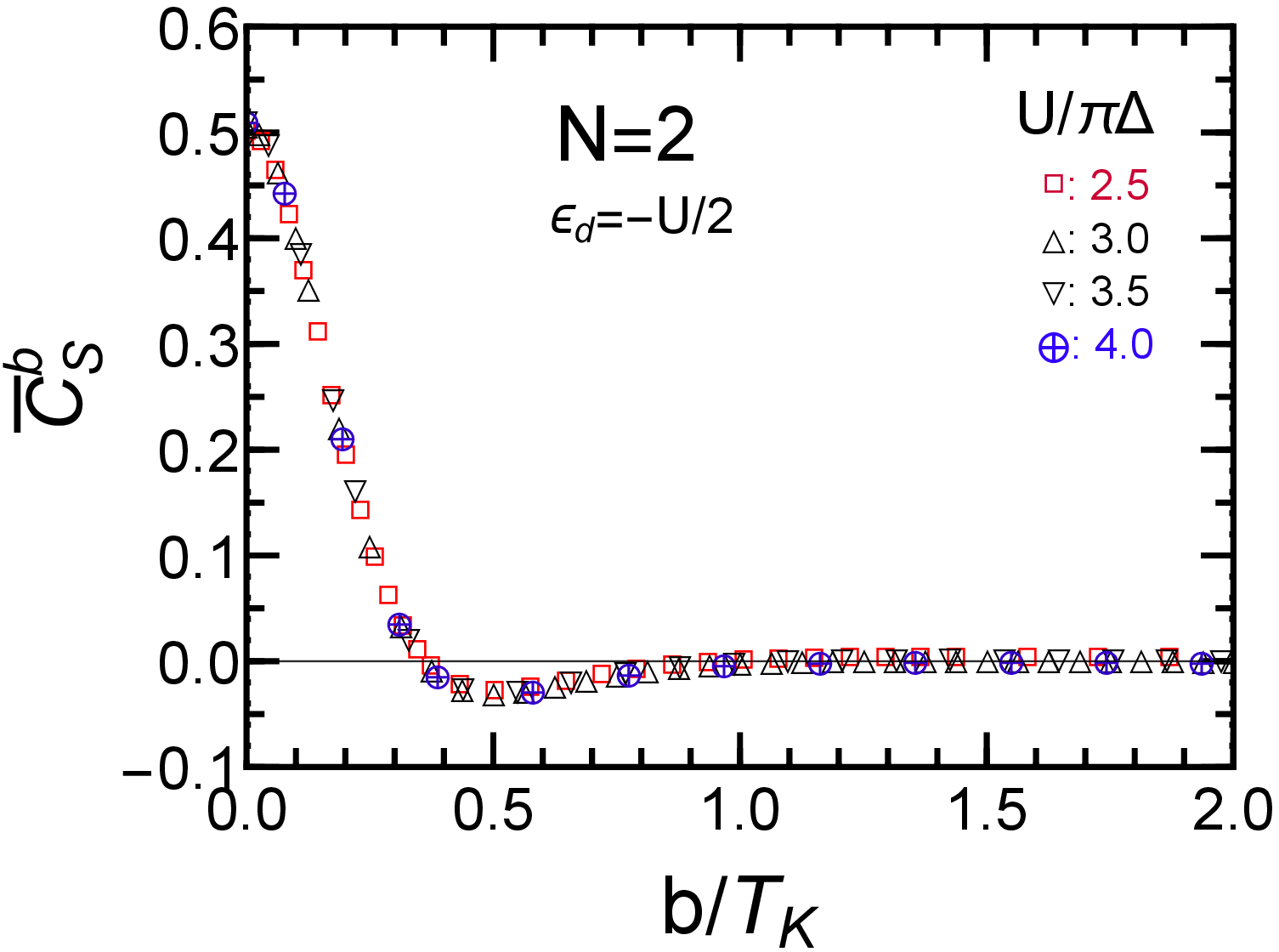}
\caption{Nonlinear current-current correlations.
Left panel:  $F_{K}^{} \equiv \frac{C_S^{}}{C_V^{}/3}$ 
vs  $\xi_{d}^{}/U$ for SU($N$) symmetric case 
for  $N=4$ ($\bullet$, $\blacklozenge$) 
and $N=6$  ($\circ$,$\lozenge$),  
for  $U/(\pi\Delta) = 1/3$ (diamonds)  and   $U/(\pi\Delta) = 5$ (circles).  
Right panel:  
$\overline{C}_{S}^{b}$   vs  $b/T_K$ for $N=2$ 
 at half filling $\epsilon_{d}^{}=-U/2$, 
for  $U/(\pi\Delta) = 2.5, 3.0, 3.5, 4.0$, with $b$  
the magnetic field and  $T_K$ the Kondo temperature at $b=0$. 
}
\label{fig:Fano_SUN_new}
\end{figure}

{\it Nonequilibrium FL fluctuations.---}
We show in the following how 
the transport coefficients  
evolve as  $N_d^{}$ varies continuously.  
The NRG results are also  plotted in 
Fig.\ \ref{fig:W_Theta_SUN}.   
The difference between the $C$'s  
near half filling  $|\xi_d| \lesssim  \frac{U}{2}$ 
is caused by the  two-body contributions $W$'s 
as the $\Theta$'s almost vanish.       
 In particular, 
 the  $T^2$ conductance 
 $C_{T}^{}$ is determined  by $W_{T}^{}$ 
over the wide filling range  $1 \lesssim N_d^{} \lesssim N-1$   
as the three-body contributions almost cancel out   
 $\Theta_\mathrm{I}^{} + (N-1)\,\Theta_\mathrm{II}^{} \approx 0$,  
reflecting the suppression of 
$\partial \chi_c^{}/\partial \epsilon_d^{}$ 
and $\partial \chi_s^{}/\partial \epsilon_d^{}$ mentioned above.  
For the thermal conductivity, 
the three-body contributions become negative in this region,   
 $\Theta_\mathrm{I}^{} 
+ \frac{5}{21}(N-1)\,\Theta_\mathrm{II}^{} \approx  
-\frac{16}{21}\,\widetilde{\Theta}_\mathrm{II}^{}$,  
but otherwise  $C_{\kappa}^\mathrm{QD}$ 
shows a similar  $\xi_d^{}$ dependence  to that of $C_{T}^{}$.     
The three-body contributions on  
the  $(eV)^2$ conductance  $C_{V}^{}$ are given by  
  $\Theta_{V}^{} \equiv \Theta_\mathrm{I}^{}+3 (N-1)\,\Theta_\mathrm{II}^{}$,  
which takes a value 
 $\Theta_{V}^{} \approx  2\,\widetilde{\Theta}_\mathrm{II}^{}$ 
in the same filling range.  
Thus,  $C_{V}^{}$ is significantly enhanced at  $N_d^{} \simeq 1$ and $N-1$ 
 where  $-\widetilde{\Theta}_\mathrm{II}$ shows a deep valley.  
It pushes  the tail of the $C_{V}^{}$ curve outside than that of  
the $C_{T}^{}$ 
 in the valence fluctuation region towards the empty or fully-occupied limit.

The $|eV|^3$ current noise also exhibits the Kondo plateau structures as shown in   
 Fig.\ \ref{fig:W_Theta_SUN}. 
For  $C_{S}^{}$,  the three-body contributions enter through  
 $\Theta_{V}^{}$ with a sinusoidal factor: 
 $-\Theta_{V}^{}  \cos 2\delta 
\approx 
\frac{\sin 4\delta}{2\pi}\,
\frac{\chi_{\sigma\sigma\sigma}^{[3]} }{\chi_{\sigma\sigma}^2}$   
over the range of  $1 \lesssim N_d^{} \lesssim N-1$.    
In the valence fluctuation regions mentioned above,
 $C_{S}^{}$  has  a minimum  caused by   
the higher-harmonic  $\sin 4 \delta$  and  $\cos 4 \delta$ contributions. 
We also find that  $C_{S}^{}$ approaches zero 
almost simultaneously with  $C_{V}^{}$ 
at $|\xi_d^{}| \simeq 1.4 U$ for $N=4$, 
and at $|\xi_d^{}| \simeq 2.4 U$ for  $N=6$.    
This proximity of the zero points  
 affects the behavior of an extended Fano factor $F_K^{}$, 
defined as  the ratio of order $(eV)^3$ current noise 
 to the nonlinear current 
\cite{MoraMocaVonDelftZarand,MoraSUnKondoII,Note2}:   
\begin{align}
\!\! 
F_K^{} 
\equiv   
\lim_{|eV| \to 0} 
\frac{ S_\mathrm{noise}^\mathrm{QD} 
-  \frac{2Ne^2 |eV|}{h}\, \frac{\sin^2  2\delta}{4} }
{- 2 |e| \, \left( J  - \frac{Ne^2|V| }{h}\, \sin^2  \delta \right)} 
\, = \, 
\frac{C_S^{}}{C_V^{}/3}
.   
\label{eq:Fano_SUN_Lett}
\end{align}
This  formula for the SU($N$) Anderson model 
includes  the result of Mora {\it et al\/},  obtained 
for $N=2$ at zero magnetic field \cite{MoraMocaVonDelftZarand},  as a special case. 
In the strong coupling limit at integer $N_d^{}$,  
it also agrees  with another noise  formula of Mora {\it et al\/}   
for the SU($N$) Kondo model \cite{MoraSUnKondoII}.

The Fano factor for $N= 4, 6$ is plotted vs $\xi_d^{}$ 
for two different values of $U$ in the left panel of Fig.\ \ref{fig:Fano_SUN_new}.   
It  reaches the local maximum 
 $F_K^{} \to  (N-1+9\widetilde{K}^2)/(N-1+5\widetilde{K}^2)$ 
at $\xi_d^{}=0$ \cite{SakanoFujiiOguri}, and 
has positive plateaus for large $U$ at integer  $N_d^{}$.
In the limit of  $|\xi_d^{}| \to \infty$,   
the ratio becomes negative and takes  the noninteracting value $F_K^{}\to -1$.    
By definition,  $F_K^{}$  changes sign at the zero points of $C_S^{}$. 
It also diverges at the zero point of $C_V^{}$,  
where the nonlinear component of $J$ 
changes direction  from backward to forward. 
Such a singularity already exists  
 for $U=0$   at  $|\xi_d^{}|={\Delta}/{\sqrt{3}}$.  
For large $U$,   $F_K^{}$ diverges near  $|\xi_d^{}| \simeq \frac{N-1}{2}U$ 
in the valence fluctuation region towards the  empty or fully-occupied limit. 
We can see that sign of the coefficient $C_S^{}$ 
at the singular points becomes positive for large  $U$, whereas  
it is negative for small $U$. 
Sign change occurs, for both $N=4$ and $6$, at a finite $U$ 
between the two examined cases  $U/(\pi\Delta)=1/3$ and  $5$. 
It is associated with the large enhancement of three-body contributions  $\Theta_{V}^{}$ 
occurring in the Kondo regime at $N_d^{} \simeq 1$ and $N-1$ for $N>2$. 
In contrast, the NRG calculations examined so far 
indicate that sign is always negative  in the SU(2)  case 
for any $U \geq 0$ \cite{MoraMocaVonDelftZarand,Note2}. 
The main difference is that   in the SU(2) case  the three-body correlations  
evolve  in the valence fluctuation region 
where  electron correlations become less important and  
the two-body contributions $W_S^{}$ dominate $C_S^{}$ 
near the singular point.

{\it Magnetic-field  dependence.---} 
We next consider effects of a magnetic field $b$ that 
breaks the SU($N$) and TR symmetries: 
specifically at half filling for $N=2$, where 
 $\epsilon_{d\uparrow}^{} = -\frac{U}{2} - b$,   
$\epsilon_{d\downarrow}^{} = -\frac{U}{2} + b$, 
and the electron filling is fixed at 
 $\langle  n_{d\uparrow} \rangle  + \langle n_{d\downarrow}\rangle =1$. 
 In this case,  
the transport coefficients can be described also by five FL  parameters: 
magnetization $m_d  \equiv  \langle  n_{d\uparrow} \rangle 
- \langle n_{d\downarrow}\rangle$,
susceptibilities  
$\chi_{\uparrow\uparrow}^{}$ $=\chi_{\downarrow\downarrow}^{}$ 
and $\chi_{\uparrow\downarrow}^{}= \chi_{\downarrow\uparrow}^{}$, 
and three-body correlations     
$ \chi_{\uparrow\uparrow\uparrow}^{[3]}
=-\chi_{\downarrow\downarrow\downarrow}^{[3]}$
and 
$\chi_{\uparrow\downarrow\downarrow}^{[3]}
= -\chi_{\uparrow\uparrow\downarrow}^{[3]}$. 
The nonlinear current for this case  
has previously been studied 
\cite{ao2017_1_PRL,ao2017_3_PRB_addendum,FilipponeMocaWeichselbaumVonDelftMora}.
However, behavior of its fluctuations has not been clarified so far.

Here, we examine the current noise  at $T=0$
\footnote{
$
\overline{C}_s^{b}  = 
\frac{\pi^2}{192}\bigl[\,
\cos (2\pi m_d)
+  
\{ 4+5\cos (2\pi m_d)\}(R-1)^2
\\
 +  \Theta_{V}^{b}   \cos (\pi  m_{d})
\,\bigr](\frac{T_K}{T^*})^2$,
and 
$\, \Theta_{V}^{b}  \equiv  
- \frac{\sin (\pi  m_{d})}{2\pi} 
\frac{\chi_{\uparrow\uparrow\uparrow}^{[3]}
+3\chi_{\uparrow\downarrow\downarrow}^{[3]}
}{\chi_{\uparrow\uparrow}^2}$
}:    
\begin{align*}  
S_\mathrm{noise}^\mathrm{QD}
\,=\,    \frac{4e^2|eV| }{h} 
\left[\,
\frac{\sin^2 (\pi m_d^{})}{4} 
\  + \ \overline{C}_{S}^{b}  \left(\frac{eV}{T_K}\right)^2
 + \cdots
\,\right].
\end{align*}  
Note that the second term is  scaled by $T_K \equiv \left. T^*\right|_{b=0}$, 
the Kondo temperature defined at zero field. 
Thus, the coefficient  $\overline{C}_{S}^{b}$ 
includes all effects of $b$, which enter through the FL parameters. 
In the right panel of Fig.\ \ref{fig:Fano_SUN_new}, 
NRG results for   $\overline{C}_{S}^{b}$ 
are plotted as a function of $b/T_K$ for several different values of $U$. 
 We find that the nonlinear noise 
exhibits a universal behavior for $U/(\pi\Delta)\gtrsim 2.0$ 
 in a similar way that  
the nonlinear current shows \cite{ao2017_1_PRL,ao2017_3_PRB_addendum}. 
It decreases rapidly as $b$ increases for small fields,    
changes sign at  $b \approx 0.36 T_K^{}$, 
 takes a minimum at $b \approx 0.5 T_K$, 
 and then approaches zero at  $b \gtrsim T_K$. 
We note that order $T^3$ thermal conductivity 
also exhibits the scaling behavior \cite{Note2}. 
These observations reflect the fact that 
the three-body fluctuations
show the universal scaling behavior  in the Kondo regime 
 without the TR symmetry.

{\it Conclusion.---} 
Nonlinear transport through the SU($N$) Anderson impurity 
has been described  in a unified way with five FL parameters.
We have demonstrated how the FL state 
evolves as electron filling $N_d^{}$ varies, using the NRG up to $N=6$.
For strong interactions $U$,   
 not only charge fluctuations but also the derivatives of 
charge and spin susceptibilities 
are suppressed  over a wide filling range  $1\lesssim N_d^{} \lesssim N-1$.
It reduces the number of variable  FL parameters from five to three, 
and causes the Kondo plateau structures  
emerging for all the coefficients $C$'s. 
In particular, the three-body contributions 
on  $C_V^{}$ are significantly enhanced at $N_d^{} \simeq 1$ and $N-1$  for  $N>2$.  
It also  affects the behavior of  nonlinear Fano factor $F_K^{}$ 
 in the valence fluctuation region.
We have also  shown that the nonlinear current noise 
 exhibits the universal magnetic-field scaling in the Kondo regime.
The FL parameters can also be deduced from experiments  
and can be used to predict behaviors of  unmeasured observables.

 \smallskip 

We would like to thank 
K.\ Kobayashi, T.\ Hata, M.\ Ferrier, R.\ Deblock, and A.\ C.\ Hewson 
for valuable discussions.  
This work was supported by 
JSPS KAKENHI Grand Numbers  JP18J10205, JP18K03495, JP26220711, 
and  JST CREST Grant No.\ JPMJCR1876.


%

\clearpage


\begin{widetext}

\begin{center}
{\large \bf 
Fermi liquid theory for nonlinear transport through a multilevel Anderson impurity
(Supplemental Material)
\rule{0cm}{0.5cm}
}
\end{center}

\smallskip

\begin{center}
Yoshimichi Teratani$^1$,   Rui Sakano$^2$, 
and Akira Oguri$^{1,3}$  
\smallskip

{\it 
$^1$Department of Physics, Osaka City University, Sumiyoshi-ku, 
Osaka 558-8585, Japan

$^2$The Institute for Solid State Physics, 
the University of Tokyo, Kashiwa, Chiba 277-8581, Japan

$^3$NITEP, Osaka City University, Sumiyoshi-ku, 
Osaka 558-8585, Japan
}
\end{center}

\smallskip 

\section* {I. \quad  Derivation of the transport coefficients 
 $\, C_{T}^{}$,  $C_{V}^{}$ and 
 $C_{\kappa}^{\mathrm{QD}}$ 
}

We describe here outline of the derivation of  the coefficients $C$'s,
listed in table \ref{tab:C_W_SUN} in the main text.  
The steady current $J$ through the quantum dots has been calculated  
using the formula given in Eq.\ \eqref{eq:current_formula} 
with the transmission probability, defined by  
\begin{align}
\mathcal{T}_{\sigma}(\omega) \,=& \   
   \frac{-4\Gamma_L \Gamma_R}{\Gamma_L+\Gamma_R} 
\  \mathrm{Im}\,G_{\sigma}^{r}(\omega) 
\, , 
\qquad \qquad 
G_{\sigma}^{r}(\omega)
 \,=\,     
\frac{1}{\omega -\epsilon_{d\sigma}^{} +i\Delta \,
- \Sigma_{\sigma}^{r}(\omega)  } \;. 
 \label{eq:Transmission_Gr_appendix}
\end{align}
We have also calculated the thermal conductivity  $\kappa_\mathrm{QD}^{}$, 
which can be expressed in the following form at  $eV=0$, 
\begin{align}
 \kappa_\mathrm{QD}^{} =   
 \frac{1}{2\pi \hbar T}
 \left[  
 \sum_{\sigma}
  \mathcal{L}_{2,\sigma}^{\mathrm{QD}} 
 - 
 \frac{ \left(
 \sum_{\sigma}
 \mathcal{L}_{1,\sigma}^{\mathrm{QD}}\right)^2}{\sum_{\sigma}
 \mathcal{L}_{0,\sigma}^{\mathrm{QD}}} \right]  \,, 
\qquad \qquad  \quad 
 \mathcal{L}_{n,\sigma}^\mathrm{QD} \, \equiv  \,
 \int_{-\infty}^{\infty}  
 d\omega\, 
 \omega^n \  
  \mathcal{T}_{\sigma}(\omega) \,
 \left( -
 \frac{\partial f(\omega)}{\partial \omega}
 \right)  .
 \label{eq:Ln_QD}
 \end{align}
The coefficients  for the charge and heat currents,    
 $C_{T}^{}$, $C_{V}^{}$,  and  $C_{\kappa}^\mathrm{QD}$, can be deduced 
 form the  low-energy expansion of the retarded self-energy 
$\Sigma_{\sigma}^{r}(\omega)$  obtained 
up to terms of order $\omega^2$, $T^2$, and $(eV)^2$.  
 We note that, in order to determine also 
 the thermopower of quantum dots $\mathcal{S}_\mathrm{TP}^\mathrm{QD}$ 
 up to next leading order,   
 additional  terms of order $\omega^3$  and $\omega\, T^2$ of the self-energy 
 are necessary.
 This is because the leading term of $\mathcal{S}_\mathrm{TP}^\mathrm{QD}$ 
 already includes  the derivative 
   $\rho_{d\sigma}' 
  \equiv 
  \frac{\partial \rho_{d\sigma}^{}(\omega)}{\partial \omega}\big|_{\omega=0}  
  $ which describes a variation from the ground state:     
 \begin{align}
 \mathcal{S}_\mathrm{TP}^\mathrm{QD} 
 \, \equiv \, 
  \frac{-1}{|e|T} \frac{\sum_{\sigma} \mathcal{L}_{1,\sigma}^{\mathrm{QD}}}
  {\sum_{\sigma} \mathcal{L}_{0,\sigma}^{\mathrm{QD}}} 
   \  = \  -\,\frac{\pi^2}{3|e|} 
  \frac{\sum_{\sigma} \rho_{d\sigma}'}{\sum_{\sigma}\rho_{d\sigma}^{}}\  T
 \,+ \, O(T^3)\,, 
\qquad \qquad \qquad 
\rho_{d\sigma}'\,=\,    \frac{\chi_{\sigma\sigma}\, \sin2 \delta_\sigma^{}}{\Delta} 
\,. 
 \end{align}

The expansion coefficients of  $\Sigma_{\sigma}^r(\omega)$ 
can be expressed in terms of the linear and nonlinear susceptibilities,
\begin{align}
\mathrm{Im}\, \Sigma_{\sigma}^r(\omega) 
\ = &  \ -\,   \frac{\pi (N-1)}{2}\,   
 \frac{ \chi_{\sigma\sigma'}^2}{\rho_{d\sigma}^{}} \,
  \left[\,\omega^2 
 + 
 \frac{3}{4} \,(eV)^2 
 +(\pi T)^2  
   \,\right] +  \cdots \,,
\qquad \qquad \quad   (\sigma' \neq \sigma) \;,  
\label{eq:self_imaginary_SUn_suppl}
\\
\epsilon_{d}^{} + 
\mathrm{Re}\, \Sigma_{\sigma}^r(\omega) 
\, = & \  \ 
 \Delta\, \cot \delta 
 \,+ \left( 1-\widetilde{\chi}_{\sigma\sigma}^{}\right) \omega 
   + 
   \, 
\frac{1}{2} \frac{\partial \widetilde{\chi}_{\sigma\sigma}^{}} 
 {\partial \epsilon_{d\sigma}}\,  \omega^2 
 \, +  
\, \frac{N-1}{6}\, \frac{\chi_{\sigma\sigma'\sigma'}^{[3]}}{\rho_{d\sigma}^{}} 
 \left[\,
 \frac{3}{4} 
 (eV)^2  + \left( \pi T\right)^2 \, \right]  + \cdots  \;,
\label{eq:self_real_ev_mag_SUn_alpha0_suppl}
\end{align}
for  $\Gamma_L=\Gamma_R=\Delta/2$ 
and  $\mu_L=-\mu_R = eV/2$ in the SU($N$) symmetric case. 
Note that   $\widetilde{\chi}_{\sigma\sigma''}^{} \equiv 
\delta_{\sigma\sigma''} 
+ \left.
\frac{\partial  \Sigma_{\sigma}^{r}(0)}{\partial \epsilon_{d\sigma''}}
\right|_{T=eV=0}^{}$, 
 $\,\chi_{\sigma\sigma''}^{}= 
\rho_{d\sigma}^{} \widetilde{\chi}_{\sigma\sigma''}^{}$, and   
  $\rho_{d\sigma}^{} = \sin^2 \delta /(\pi \Delta)$ 
for $\sigma =1,2,3, \ldots, N$. 
These results 
are obtained by  extending further the latest version of Fermi-liquid description 
[A.\ Oguri and A.\ C.\ Hewson, Phys.\ Rev.\ B {\bf 97}, 035435] 
to the multilevel cases  $N>2$.  
At $T=eV=0$, the  self-energy satisfies the Ward identity of the following form,  
which  yields the Fermi-liquid relations between the expansion coefficients,  
\begin{align}
& 
\left(
 \delta_{\sigma\sigma'} \, 
\frac{\partial }{\partial \omega} 
\ + \   \frac{\partial }
{\partial \epsilon_{d\sigma'}^{}} 
\right) \Sigma_{\sigma}^{--}(\omega) 
\  = \ 
- 
\,
\Gamma_{\sigma\sigma';\sigma'\sigma}^{--;--}
(\omega,0;0,\omega) 
\,
\rho_{d\sigma'}^{} \;.
\label{eq:YYY_T0_causal_X} 
\end{align}
Here, 
$
\Sigma_{\sigma}^{--}(\omega) = 
\mathrm{Re}\,\Sigma_{\sigma}^{r}(\omega) 
+i\, \mathrm{sgn}(\omega)\,
\mathrm{Im}\,\Sigma_{\sigma}^{r}(\omega)$, and 
 $\Gamma_{\sigma\sigma';\sigma'\sigma}^{--;--}
(\omega, \omega'; \omega' ,\omega)$ 
is the causal vertex function at $T=eV=0$. 
It has been determined up to linear order terms with respect to $\omega$ and $\omega'$, 
\begin{align}
\Gamma_{\sigma\sigma;\sigma\sigma}^{--;--}
(\omega , \omega'; \omega', \omega) 
\rho_{d\sigma}^{2}
\, = & \   
 i \pi 
\sum_{\sigma'(\neq \sigma)}
\chi_{\sigma\sigma'}^2
\bigl|\omega-\omega' \bigr| 
\ + \  \cdots \;, 
 \qquad  \qquad \qquad \qquad   (\sigma' \neq \sigma) \;,  
 \label{eq:GammaUU_general_omega_dash_N}
\\
\Gamma_{\sigma\sigma';\sigma'\sigma}^{--;--}
(\omega, \omega'; \omega' ,\omega) 
\,\rho_{d\sigma}^{}\rho_{d\sigma'}^{}
\,  =&   \ 
 -
\chi_{\sigma\sigma'}^{}
+ 
\rho_{d\sigma}^{}
\frac{\partial \widetilde{\chi}_{\sigma\sigma'}}
{\partial \epsilon_{d\sigma}^{}} \, \omega  
+ 
\rho_{d\sigma'}^{}
\frac{\partial \widetilde{\chi}_{\sigma'\sigma}}
{\partial \epsilon_{d\sigma'}^{}} \, \omega'   
 + 
i \pi \,
\chi_{\sigma\sigma'}^2 \, 
\Bigl(
\,\bigl|  \omega - \omega'\bigr| 
-
\,\bigl| \omega + \omega' \bigr| 
\,\Bigr)
+ \cdots
\;.
\label{eq:GammaUD_general_omega_dash_N}
\end{align}
The Ward identity itself follows from the current conservation between the dot and leads:  
\begin{align}
\frac{\partial}{\partial t}\,(e\,n_{d\sigma}^{\phantom{0}})  
+ \widehat{J}_{R,\sigma} - \widehat{J}_{L,\sigma} = 0 , \qquad  \quad 
\widehat{J}_{L,\sigma} \equiv   i\,e v_L 
\left(
\psi^{\dagger}_{L\sigma} d^{}_{\sigma} 
-d^{\dagger}_{\sigma} \psi^{}_{L\sigma} \right) , 
\qquad 
\widehat{J}_{R,\sigma} \equiv -i\,e v_R 
\left(
\psi^{\dagger}_{R\sigma} d^{}_{\sigma} 
-d^{\dagger}_{\sigma} \psi^{}_{R\sigma}\right) \;. 
 \end{align}

\begin{figure}[b]
 \leavevmode
\begin{minipage}{1\linewidth}
\raisebox{0.4cm}{
\includegraphics[width=0.2\linewidth]{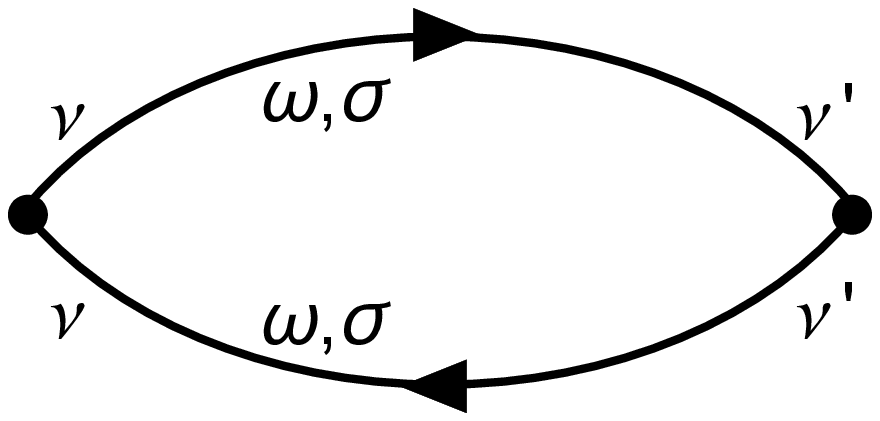}
}
\rule{0.15\linewidth}{0cm}
\includegraphics[width=0.24\linewidth]{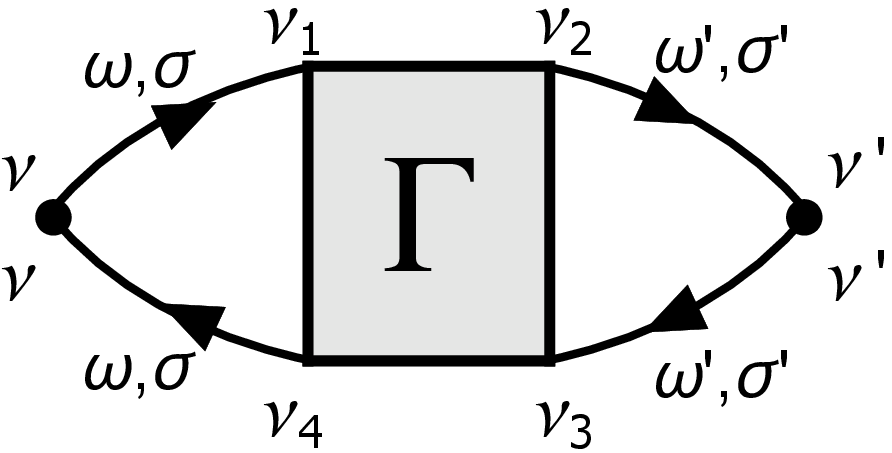}
 \caption{
The Feynman diagrams for  the 
correlation function  
  $ \int_{-\infty}^{\infty} \! dt \, 
 \mathcal{K}_{\sigma'\sigma}^{\nu'\nu}(t,0)$.
The solid lines denote 
the Keldysh Green's functions 
  $G_{\sigma}^{\nu'\nu}(\omega)$. 
The shaded region in the diagram on the right represents    
the Keldysh vertex function 
$\Gamma_{\sigma\sigma';\sigma'\sigma}^{\nu_1\nu_2;\nu_3\nu_4}
(\omega,\omega';\omega'\omega)$. 
The superscripts  $\nu$, $\nu'$  and   $\nu_i $ ($i=1,2,3,4$) 
specify  the branches of  Kedysh time-loop contour.
We are using the notation in which  $\nu=-$ and $+$ represent 
the forward and return paths, respectively.
 }
\label{fig:Kubo_Keldysh_digaram}
\end{minipage}
\end{figure}

\section* {II. \quad  Fermi-liquid corrections 
for nonlinear current noise  $\,S_\mathrm{noise}^\mathrm{QD}$ }

In contrast to the average current $J$ and 
thermal conductivity  $\kappa_\mathrm{QD}^{}$, 
the current noise depends also on the two-quasiparticle collisions 
which correspond to the vertex corrections 
for the current-current correlation function  $\mathcal{K}_{\sigma'\sigma}^{\nu'\nu}$:  
\begin{align}
S_\mathrm{noise}^\mathrm{QD}\,= &  \  
\int_{-\infty}^{\infty} \!\!  dt\  
\sum_{\sigma\sigma'}
i \Bigl[\,
\mathcal{K}_{\sigma'\sigma}^{+-}(t,0)
+\mathcal{K}_{\sigma'\sigma}^{-+}(t,0)
\,\Bigr] 
\;, 
\qquad \qquad 
\delta \widehat{J}_{\sigma}(t)  \equiv  
\widehat{J}_{\sigma}(t) - \langle  
\widehat{J}_{\sigma}(t)\rangle_{V}^{} 
\,,
\\
\mathcal{K}_{\sigma'\sigma}^{+-}(t,0)
\, \equiv  & \   
-i \,\bigl\langle \delta \widehat{J}_{\sigma'}(t) 
\, \delta \widehat{J}_{\sigma}(0) \bigr\rangle_{V}^{}\,, \qquad
\qquad  
\mathcal{K}_{\sigma'\sigma}^{-+}(t,0)
\, \equiv  \,
-i \,\bigl\langle \delta \widehat{J}_{\sigma}(0) 
\, \delta \widehat{J}_{\sigma'}(t)\bigr\rangle_{V}^{}
\;
 \rule{0cm}{0.6cm}
\,. 
\end{align}
Here,   
 $\widehat{J}_{\sigma}
 \equiv    
(
\Gamma_L \widehat{J}_{R,\sigma} 
+\Gamma_R \widehat{J}_{L,\sigma})/(\Gamma_L+\Gamma_R)
$ is a symmetrized current operator.
In this work, we have  expanded  
$S_\mathrm{noise}^\mathrm{QD}$  up to terms of order $(eV)^3$,  
 using the diagrammatic representation  illustrated 
 in Fig.\ \ref{fig:Kubo_Keldysh_digaram}.  
To this end, 
all the components of  Keldysh Green's function $G_{\sigma}^{\nu'\nu}(\omega)$ 
have been deduced up to order $\omega^2$ and $(eV)^2$, 
and the Keldysh vertex function    
$\Gamma_{\sigma\sigma';\sigma'\sigma}^{\nu_1\nu_2;\nu_3\nu_4}
(\omega,\omega';\omega'\omega)$  have been calculated up to  
linear order in  $\omega$, $\omega'$ and $eV$. 
We have checked that the result satisfies 
the  nonequilibrium Ward identity 
 [A.\ Oguri, Y.\ Tetatani, and S.\ Sakano, unpublished], 
\begin{align}
 \left(
\delta_{\sigma\sigma' }\,  \frac{\partial}{\partial \omega} 
+ \frac{\partial}{\partial \epsilon_{d\sigma'}}
\right)
\Sigma_{\sigma}^{\nu_4\nu_1}(\omega)
\, = \,  
-\int_{-\infty}^{\infty}\! \frac{d \omega'}{2\pi}\, 
\sum_{\nu_2\nu_3}\,
\Gamma_{\sigma\sigma';\sigma'\sigma}^{\nu_1\nu_2;\nu_3\nu_4}
 (\omega,\omega'; \omega',\omega)
\  2 \Delta
\,
G_{\sigma'}^{r}(\omega') \,G_{\sigma'}^{a}(\omega')
\,
\left(
-\frac{\partial  f_\mathrm{eff}^{} (\omega')}{\partial \omega'}
\right) 
\,.
\label{eq:WT_AO} 
\end{align}
Here,  
 $f_{\mathrm{eff}}(\omega) 
= \bigl\{ \Gamma_L\,f_L(\omega) + \Gamma_R\,f_R(\omega)\bigr\} 
/(\Gamma_L +\Gamma_R)$,  
and  $\Sigma_{\sigma}^{\nu_4\nu_1}(\omega)$ is the Keldysh self-energy. 
The expansion coefficient for order $(eV)^3$  current noise   
is given in  table \ref{tab:C_W_SUN}.  
It can be separated into 
two parts $C_S^{} = C_S^\mathrm{qp} +  C_S^\mathrm{coll}$,
as mentioned  in the main text.
Here, $C_S^\mathrm{qp}$ and $C_S^\mathrm{coll}$ 
represent respectively 
 the  contributions of  the bubble diagram and that of  the vertex corrections  
shown in Fig.\ \ref{fig:Kubo_Keldysh_digaram}. 


The nonlinear Fano factor,  $F_K=\frac{C_S^{}}{C_V^{}/3}$,  
can be expressed in the following form 
for the SU($N$) Anderson impurity at arbitrary electron fillings,        
\begin{align}
F_K^{} \,= \, 
\frac{\cos 4 \delta\,
+ 
\Bigl[\,
 4+5\cos 4 \delta  + 
\frac{3}{2}\bigl(1- \cos 4\delta\bigr)\,(N-2)
\,\Bigr] \, \displaystyle \frac{\widetilde{K}^2}{N-1}
\,-\,
 \cos 2\delta\,\Bigl[\,
  \Theta_\mathrm{I}^{}
+
3\, (N-1) \Theta_\mathrm{II}^{}
\,\Bigr] 
}{
-
\left[1+ 5\,\displaystyle 
\frac{\widetilde{K}^2}{N-1}
\right]
\cos 2 \delta 
\,+\, 
  \Theta_\mathrm{I}^{}
 +
 3 \,(N-1) \Theta_\mathrm{II}^{}
}
\;.
\label{eq:FK_SUN}
\end{align}
For $N=2$, it reproduces the previous result, obtained by Mora {\it et al\/}  
[Eq.\ (11) of  Phys.\ Rev.\ B {\bf 92}, 075120 (2015)]: 
their notation and our one correspond to each other such that 
$
\alpha_{\sigma}^{(1)}/\pi=    
  \chi_{\sigma\sigma}^{}$,
$
\phi_{\sigma\sigma'}^{(1)}/\pi= 
 - \chi_{\sigma\sigma'}^{}$, 
$\alpha_{\sigma}^{(2)}/\pi 
 =  
-\frac{1}{2}\, 
\chi_{\sigma\sigma\sigma}^{[3]}$, and 
$\phi_{\sigma\sigma'}^{(2)}/\pi 
=
 2 \, 
\chi_{\sigma\sigma'\sigma'}^{[3]}
$ for $\sigma'\neq \sigma$.
In Fig.\ \ref{fig:FL_parameters_SU2},  
we have also  plotted  $F_K^{}$ 
and other Fermi-liquid parameters for the SU(2) symmetric case as functions 
of $\epsilon_d^{}$ for comparisons 
with those for $N=4$ and $6$ shown in the main text.

Equation \eqref{eq:FK_SUN} also reproduces the previous result 
in the particle-hole symmetric case,  
at which  $\delta = \pi/2$ and the three-body contributions 
vanish $\Theta_\mathrm{I}=\Theta_\mathrm{II}=0$ 
[Sakano {\it et al\/} Phys.\ Rev.\ B {\bf 83} , 075440 (2011)] :
\begin{align}
F_K^{} 
\,\xrightarrow{\, \epsilon_d  \to -(N-1)U/2\,}\,
\frac{1+\frac{9\widetilde{K}^2}{N-1}}{1+\frac{5\widetilde{K}^2}{N-1}} 
\,\xrightarrow{\, U \to \infty\,}\,
 \frac{1+\frac{9}{N-1} }{1+\frac{5}{N-1}}  
\,. 
\label{eq:FK_SUN_half_filling}
\end{align}
In the strong coupling limit $U\to \infty$, 
the occupation number  $N_d^{} \equiv \sum_\sigma \langle n_{d\sigma}^{}\rangle$ 
 becomes integer $M=1,\,2,\,\ldots,N-1$ at  $\epsilon_d^{}= -(M-1/2)U$ and 
the phase shift is locked at $\delta = \pi M/N$. 
The charge and spin susceptibilities satisfy the stationary conditions in this case, 
and thus  Eq.\ \eqref{eq:FK_SUN} can be rewritten as  
\begin{align}
\!\!\!\!\!\!\!\!\!\!\!\!\!\!\!\!\!\!\!\!
F_K^{} \  \to &  \ 
 \frac{
1+\sin^2 \left(\frac{2\pi M}{N}\right) \,
 + 
 \frac{9-13\sin^2 \left(\frac{2\pi M}{N}\right)}{N-1} 
\ + \ 2\,
\Theta_\mathrm{I}^{}\,\cos \left(\frac{2\pi M}{N}\right) 
}{
-
\left[ \, 1+ \frac{5}{N-1}\,\right]
\cos \left(\frac{2\pi M}{N}\right)
-2  \, \Theta_\mathrm{I}^{}
} \;.
\end{align}
This  expression is consistent with the corresponding noise formula 
for the SU($N$) Kondo model, obtained by Mora {\it et al\/}  
  [Eq.\ (51) of Phys.\ Rev.\ B {\bf 80}, 155322 (2009),  
after inserting some parenthesis for correcting minor typos].

\begin{figure}[t]

 \leavevmode
 \centering
\includegraphics[width=0.22\linewidth]{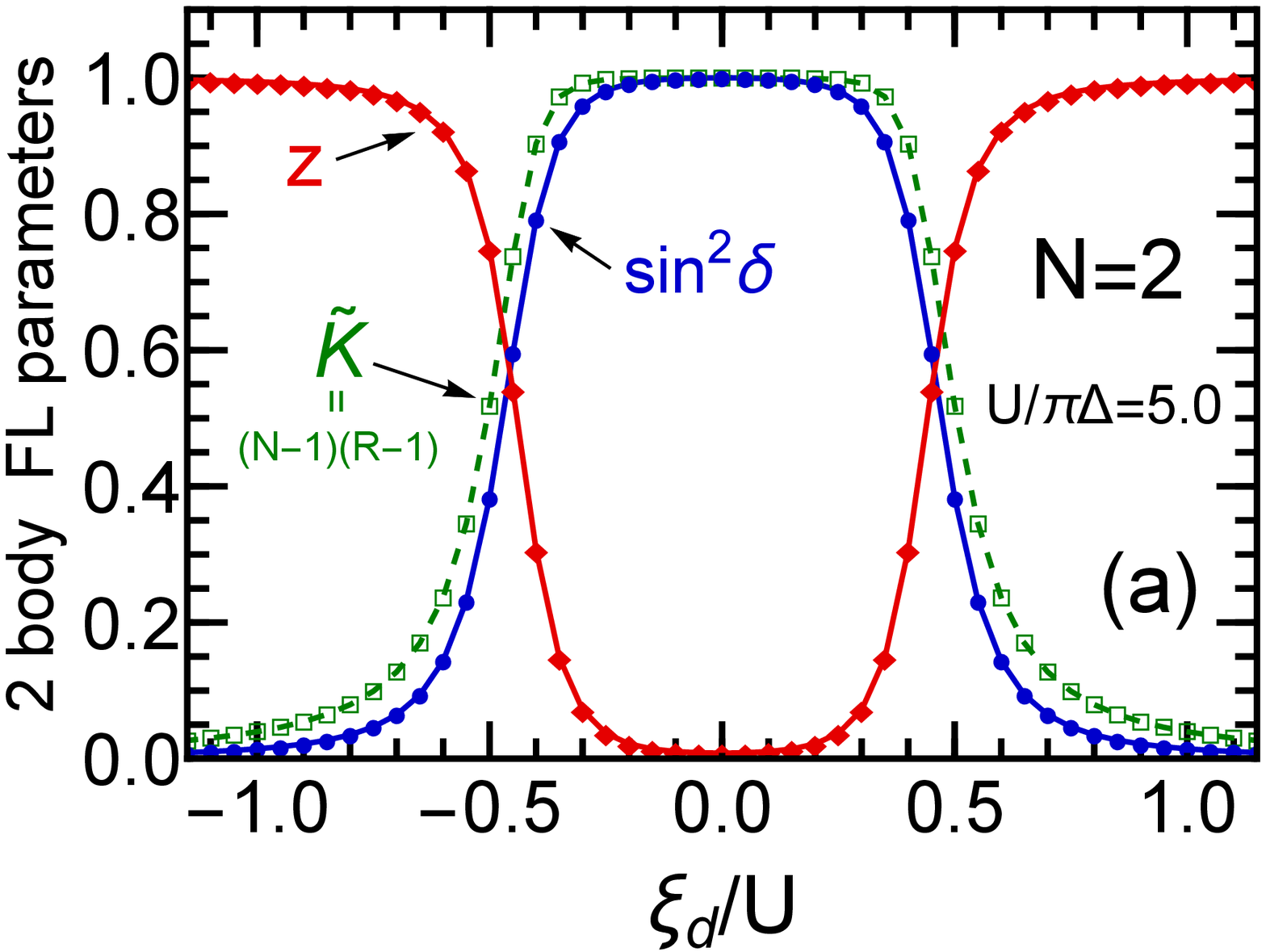}
 \hspace{0.02\linewidth}
\includegraphics[width=0.22\linewidth]{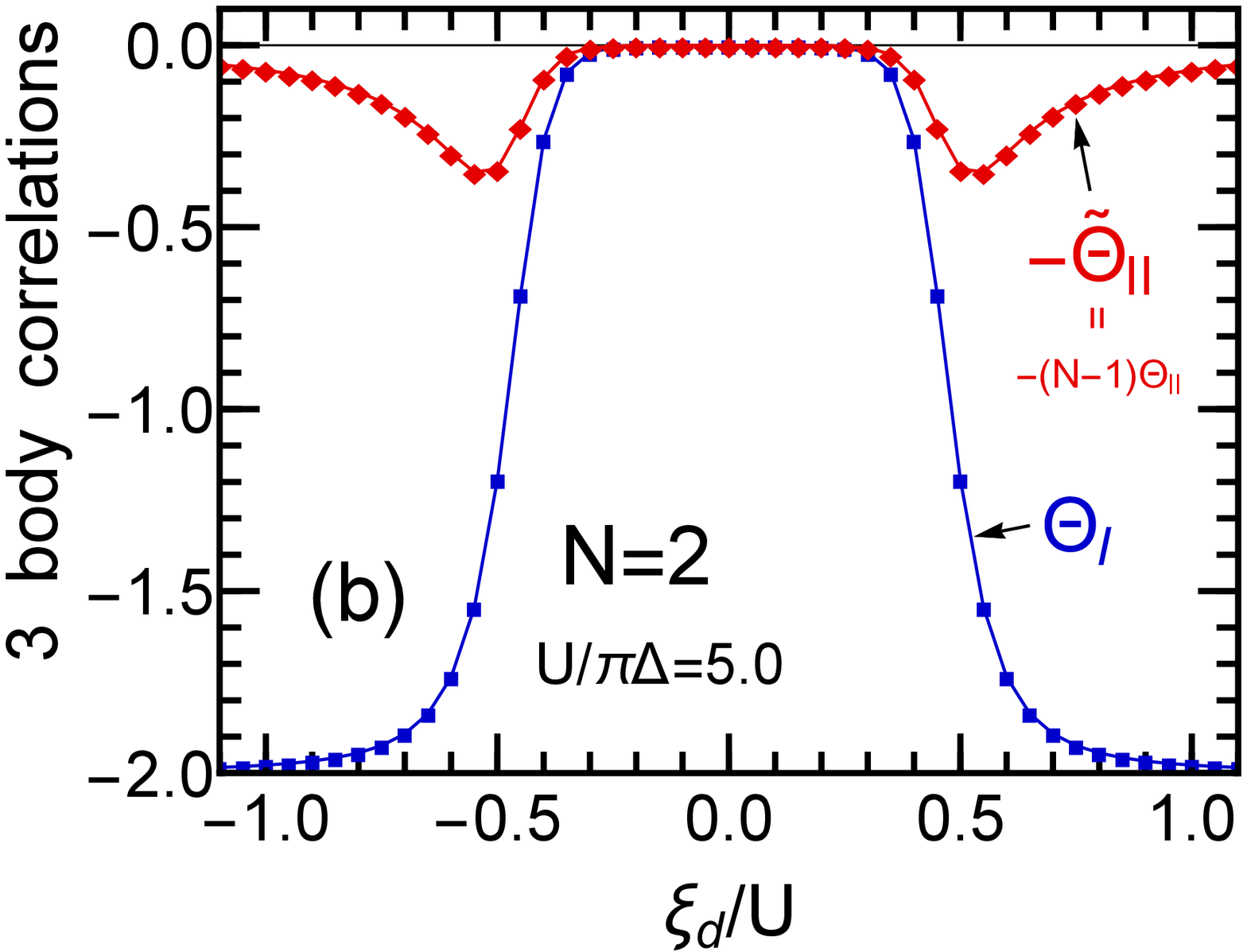}
 \hspace{0.02\linewidth}
\includegraphics[width=0.22\linewidth]{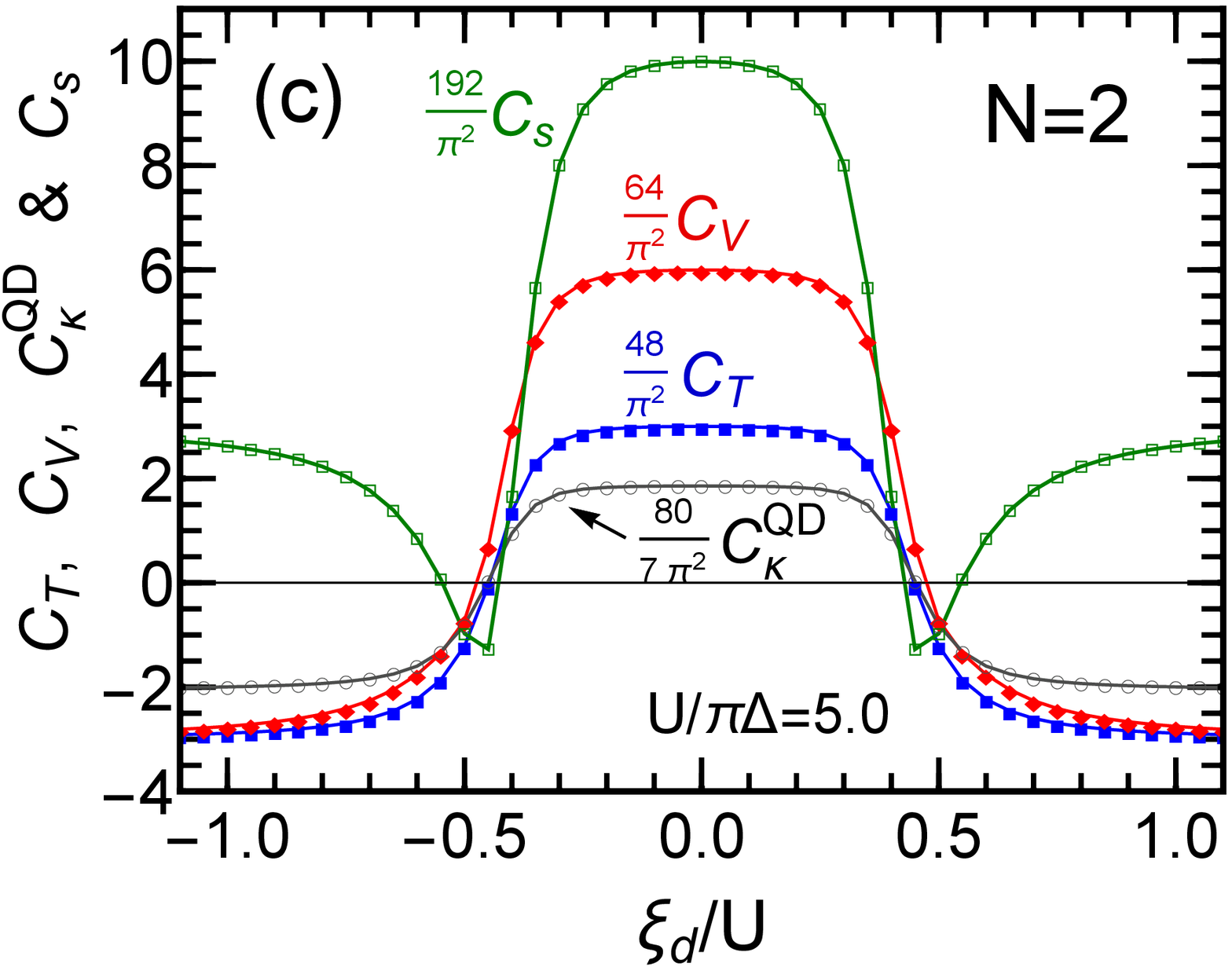}
 \hspace{0.02\linewidth}
\includegraphics[width=0.22\linewidth]{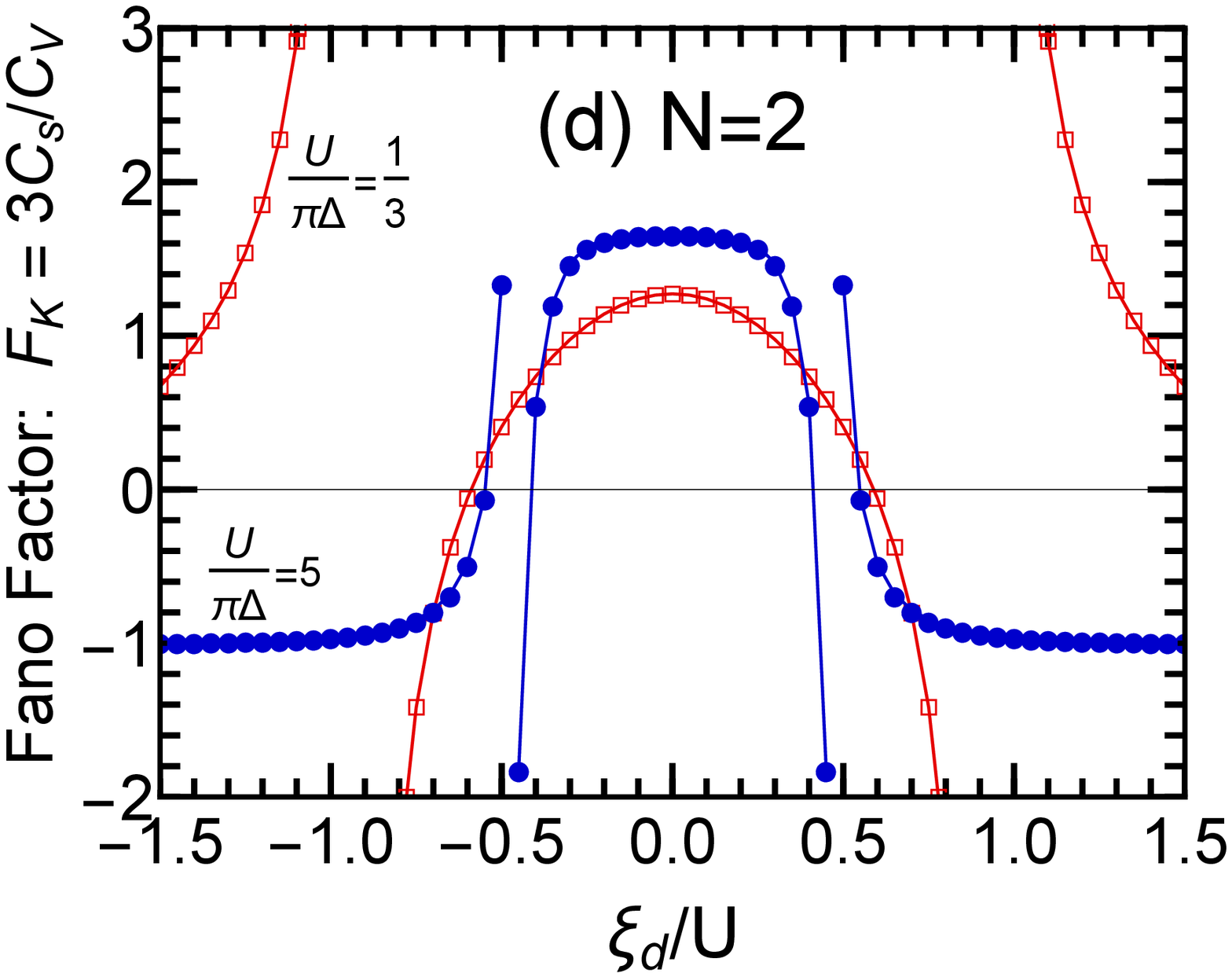}
\caption{ 
Fermi-liquid parameters for the SU(2) 
symmetric case  are plotted  
vs  $\xi_{d}^{} \equiv \epsilon_d +U/2$
for $U/(\pi\Delta)=5.0$. 
(a):    
 $\sin^2 \delta$,  renormalization factor $z$,  and 
 $\widetilde{K}\equiv (N-1)(R-1)$. 
(b):  thee-body correlatins  $\Theta_\mathrm{I}^{}$, and 
$\,\,-\widetilde{\Theta}_\mathrm{II}^{} 
\equiv  -(N-1) \Theta_\mathrm{II}^{}$.
(c): $\frac{48}{\pi^2} C_{T}^{}$,
$\frac{64}{\pi^2} C_{V}^{} $,
 $\frac{192}{\pi^2} C_{S}^{}$, and 
$\frac{80}{7\pi^2}C_{\kappa}^{\mathrm{QD}}$.
(d): nonlinear Fano factor $F_K=\frac{C_S^{}}{C_V^{}/3}$ 
is plotted also for $U/(\pi\Delta)=1/3, \, 5$. 
}
\label{fig:FL_parameters_SU2}
\end{figure}

\section*{III. \quad NRG calculations}

NRG calculations for the SU($N$) Anderson model  for $N=2, 4, 6$ 
have been carried out,  dividing $N$ channels into $N/2$ pairs     
and  exploiting the SU(2) spin and U(1) charge symmetries for each of the pairs, 
i.e.\  using  $\prod_{k=1}^{\frac{N}{2}}
\left\{\mbox{SU(2)}\otimes \mbox{U(1)}\right\}_k$ symmetries. 
The discretization parameter $\Lambda$ 
and  the number of retained low-lying excited states $N_\mathrm{trunc}$  
are chosen such that  $ (\Lambda,N_\mathrm{trunc})=
(2,4000)$ for $N=2$,  $\, (6,10000)$ for $N=4$, 
and $\, (20,30000)$ for $N=6$. 
We have also exploited methods of Stadler's {\it et al\/}
[Phys.\ Rev.\ B {\bf 93}, 235101 (2017)] for $N = 6$.
 The truncation is performed  at each step 
after adding states from each pair of the channels,  
using  Olivera's  $\mathcal{Z}$-trick [Phys.\ Rev.\ B {\bf 49}, 11986 (1994)] and 
choosing different $\mathcal{Z}$ values for different pairs:  
$\mathcal{Z}_i^{}=  1/2 +i/N $ for the $i$-th pair ($i=1,2,\ldots, N/2$).

\begin{table}[h]
\caption{The coefficients  $C$'s at finite magnetic fields $b$   
 for $N=2$ at half filling $\epsilon_d^{}=-\frac{U}{2}$.}  
 See also, A.\ Oguri and A.\ C.\ Hewson, Phys.\ Rev.\ B {\bf 98}, 079905 (E).
\begin{tabular}{l|l} 
\hline \hline
\ $C_{S}^{b} 
\ =  \,   
\frac{\pi^2}{192} 
\left[\, W_S^{b}  
+
\left(
\Theta_\mathrm{I}^{b}
+3 \Theta_\mathrm{II}^{b}
\right)  \, \cos (\pi  m_{d})
\,\right]$ \ \ \ \ \  \ \ 
& \ \ \ 
$
W_S^{b} 
 \,   \equiv  \, 
\cos  (2\pi m_d^{}) 
+
\bigl[\,4+5 \cos  (2\pi m_d^{}) \,\bigr] (R-1)^2
 $ 
\rule{0cm}{0.45cm}
\\
\ $
C_{V}^{b} 
\,=  \, 
\frac{\pi^2}{64}
 \,\bigl[
\,
W_{V}^{b} 
\,+ \,
\Theta_\mathrm{I}^{b} 
+
3\,
\Theta_\mathrm{II}^{b} 
\, \bigr]$ 
& \  \ \ 
$
W_{V}^{b} \,\equiv \, 
\left[ \,1+  5\left(R-1\right)^2 \,\right]\cos  (\pi m_d) 
 $ 
\rule{0cm}{0.45cm}
\\
\ $
C_{T}^{b} 
\,=  \,  
\frac{\pi^2}{48} 
\,\bigl[\,
W_{T}^{b} 
\,+\, 
\Theta_\mathrm{I}^{b} 
+
\Theta_\mathrm{II}^{b} 
\,\bigr]
 \quad$
& \ \ \  
$
W_{T}^{b} 
\ \equiv \, 
\left[\,1+  2\left(R-1\right)^2 \,\right]\cos  (\pi m_d)
$
\rule{0cm}{0.45cm}
\\
\ $
C_{\kappa,b}^\mathrm{QD} 
=  \,  
\frac{7\pi^2}{80} 
\,\bigl[\,
W_{\kappa,b}^\mathrm{QD} 
\,+\, 
\Theta_\mathrm{I}^{b} 
+
\frac{5}{21}
\Theta_\mathrm{II}^{b} 
\,\bigr]
$ 
  &  \ \ \  
$ 
W_{\kappa,b}^\mathrm{QD} 
\, \equiv  \, 
\left[\, 1
 + \frac{6}{7}
 \left( R -1 \right)^2 \,\right]  \cos  \left(\pi m_d\right) 
$ 
\rule{0cm}{0.45cm}
\\
\hline
\hline
\end{tabular}
\label{tab:C_and_W_N2_mag_half-filling_QD}
\end{table}

\clearpage

\section*{IV. \quad Magnetic-field dependence of current noise for $N=2$}

\end{widetext}

We describe here supplemental information about 
the nonlinear current noise at finite magnetic field $b$,  
specifically for $N=2$  at half filling,  
 where the impurity level  is given by   
 $\epsilon_{d\sigma}^{} \equiv -\frac{U}{2} - \mathrm{sgn}(\sigma)\, b$ with     
 $\mathrm{sgn}(\uparrow)= +1$  and  $\mathrm{sgn}(\downarrow)= -1$.
In this case, 
the phase shift takes the form 
$\delta_{\sigma}^{} = \pi \{1 + \mathrm{sgn}(\sigma)\, m_{d}\} /2$ 
with  $m_d  \equiv \langle  n_{d\uparrow} \rangle 
- \langle n_{d\downarrow}\rangle$,
and the other correlation functions have symmetry properties:  
$\,\chi_{\uparrow\uparrow}^{}  =  \chi_{\downarrow\downarrow}^{}$,  
$\,\chi_{\uparrow\downarrow}^{}  =  \chi_{\downarrow\uparrow}^{}$,  
 $\,\chi_{\uparrow\downarrow\downarrow}^{[3]} 
= -\chi_{\uparrow\uparrow\downarrow}^{[3]}$, and
$\,\chi_{\downarrow\downarrow\downarrow}^{[3]} 
= -\chi_{\uparrow\uparrow\uparrow}^{[3]}$. 
 Thus, the transport coefficients up to the next leading order 
can be described by five parameters, for instance, $m_d^{}$, 
 $T^* = 1/(4\chi_{\uparrow\uparrow}^{})$,
$R_{}^{}  = 1-\chi_{\uparrow\downarrow}^{}/ \chi_{\uparrow\uparrow}^{}$ 
and the following 3-body correlation functions, 
\begin{align}
\!\!\! 
\Theta_\mathrm{I}^{b}
\equiv 
- \frac{\sin (\pi  m_{d}) }{2\pi } 
\,\frac{\chi_{\uparrow\uparrow\uparrow}^{[3]}}{\chi_{\uparrow\uparrow}^2}
, \quad \  
\Theta_\mathrm{II}^{b} 
\equiv  
 - \frac{\sin (\pi  m_{d}) }{2\pi} 
\frac{\chi_{\uparrow\downarrow\downarrow}^{[3]}}{\chi_{\uparrow\uparrow}^2}.
\end{align}

The low-energy expansion of  the current noise 
  $S_\mathrm{noise}^\mathrm{QD}$,   
conductance $dJ/dV$,  
and thermal conductivity $\kappa_\mathrm{QD}^{}$ 
 for this case can be written in the following form,   
with the coefficients $C$'s  listed in  
table \ref{tab:C_and_W_N2_mag_half-filling_QD}, 
\begin{align}
&
\!\!
S_\mathrm{noise}^\mathrm{QD}  
 =  
2 \,\frac{2e^2}{h}   |eV| \! 
\left[ 
\frac{\sin^2 (\pi m_d^{})}{4} 
   +  
C_S^{b}  \left( \frac{eV}{T^*}\right)^2 
\!\!  +   \cdots  
  \right] \!  ,  \! 
\\
&
\!\!
\frac{dJ}{dV}  =   
\frac{2e^2}{h} \! 
\left[
\cos^2
 \left( \frac{\pi m_{d}}{2} \right)  
- C_{T}^{b} 
\left(\frac{\pi T}{T^*}\right)^2 \!
-   C_{V}^{b} 
 \left(\frac{eV}{T^*} \right)^2  \!\!  +  \cdots 
\right] \!   ,
\nonumber 
\\
&
\!\!
\kappa_\mathrm{QD}^{} = 
\frac{2\pi^2 T}{3h}
  \left[
\cos^2 
 \left( \frac{\pi m_{d}}{2} \right)  
- 
 C_{\kappa,b}^\mathrm{QD} 
\left( \frac{\pi T}{T^*}\right)^2
\!+ \cdots
\right] .
\nonumber 
\end{align}

\pagebreak

\begin{figure}[h]
 \leavevmode
\centering
\includegraphics[width=0.48\linewidth]{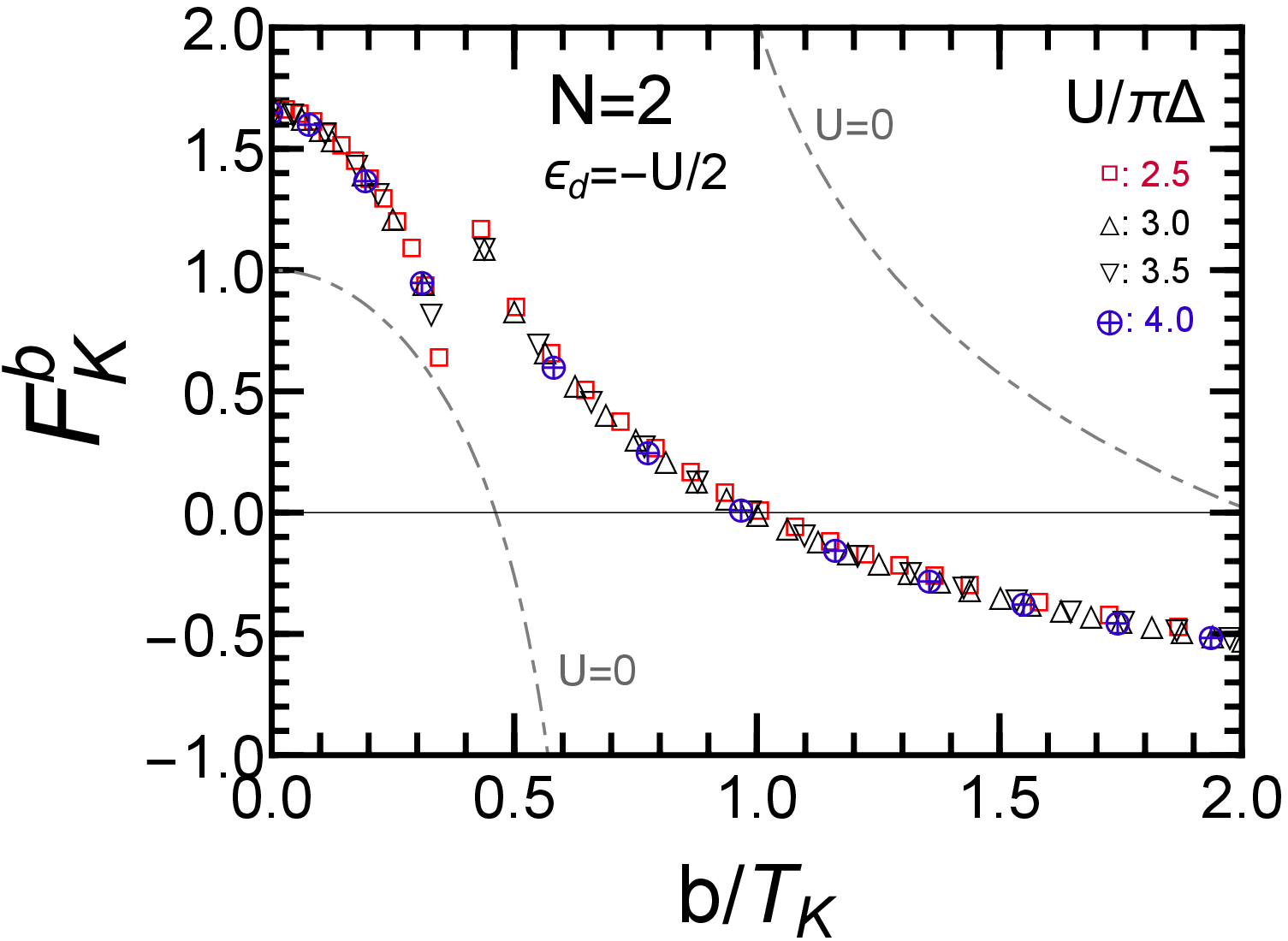}
\includegraphics[width=0.48\linewidth]{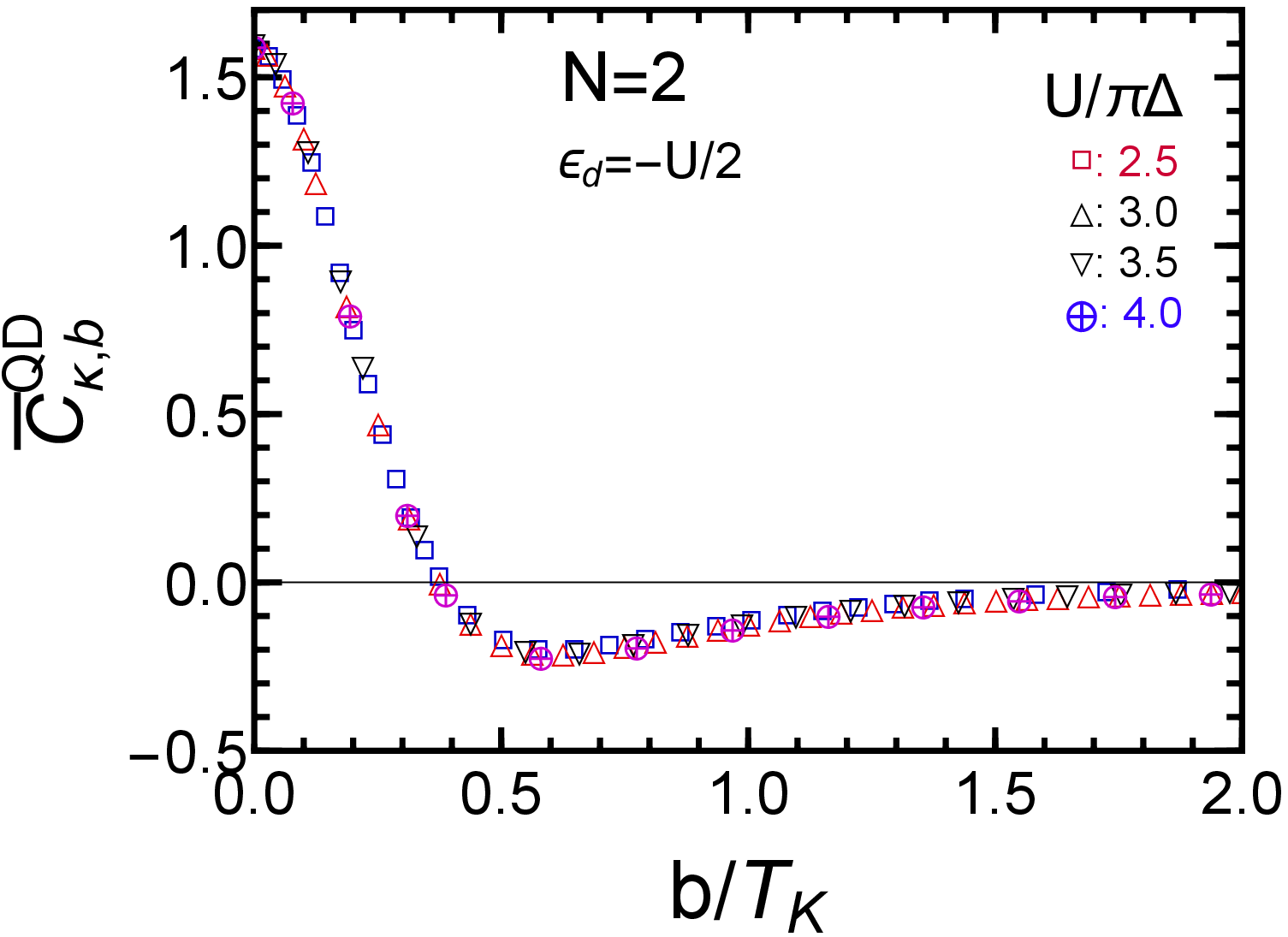}
\caption{
 $F_K^b=\frac{\overline{C}_{S}^{b}}{\overline{C}_{V}^{b}/3}$ 
and 
 $\overline{C}_{\kappa,b}^\mathrm{QD}  
\equiv (T_K/T^*)^2  C_{\kappa,b}^\mathrm{QD}$ for $N=2$ 
are  plotted  vs  $b/T_K$  at half filling     
for  $U/(\pi\Delta) = 2.5, 3.0, 3.5, 4.0$:
 $T_K$ varies with $U$, and $T_K \xrightarrow{U \to 0\,} \pi \Delta/4$.
}
\label{fig:Fano_Factor_spin_half_mag_suppl}
\end{figure}

In order to see the magnetic field dependences in the Kondo regime, 
it is preferable to  rescale the next leading $(eV)^2$ and $T^2$ contributions  
by the Kondo temperature defined at zero field  $T_K = \lim_{b\to 0} T^*$.
This is because  all effects of $b$ are absorbed into the coefficients redefined such that   
$\,\overline{C}_{V}^{b} 
\equiv (T_K/T^*)^2 \,C_{V}^{b}$,  
$\,\overline{C}_{S}^{b} \equiv(T_K/T^*)^2  \,C_{S}^{b}$, 
and 
$\overline{C}_{\kappa,b}^\mathrm{QD} 
\equiv  (T_K/T^*)^2 \,C_{\kappa,b}^\mathrm{QD}$.   
We have presented  the NRG results for 
the nonlinear current noise $\overline{C}_{S}^{b}$  in the main text;
  $\overline{C}_{V}^{b}$ was examined previously 
 [A.\ Oguri and A.\ C.\ Hewson, Phys.\ Rev.\ Lett.\ {\bf 120}, 126802 (2018)].  
In Fig.\ \ref{fig:Fano_Factor_spin_half_mag_suppl}, 
 $F_K^{b} \equiv \frac{C_{S}^{b}}{C_{V}^{b}/3}$ 
and $\overline{C}_{\kappa,b}^\mathrm{QD}$ are  plotted 
as functions of  $b/T_K$ for several different values of $U$.
The nonlinear Fano factor $F_K^{b}$  shows 
the Kondo scaling behavior for strong interactions $U/(\pi\Delta)\gtrsim 2.0$. 
 The universal curve of $F_K^{b}$ deviates significantly from the curve for $U=0$  
keeping its qualitative characteristics unchanged. 
We also find that the thermal conductivity 
$\overline{C}_{\kappa,b}^\mathrm{QD}$ 
 exhibits  the universal scaling behavior.


\end{document}